\documentclass[a4paper,11pt]{article}
\UseRawInputEncoding
\usepackage{jheppub}
\usepackage{float}
\usepackage{amsmath,amssymb}
\usepackage{bm}
\usepackage{slashed}
\usepackage{epsfig}
\usepackage{graphicx}
\usepackage{hyperref}
\usepackage{xcolor}
\usepackage{braket}
\usepackage{subfigure}
\usepackage{soul}

\hypersetup{
    colorlinks=true,
    linkcolor=blue,
    filecolor=magenta,      
    urlcolor=blue,
    citecolor=blue
}
\begin{document}
\title{Quantum computation in fermionic thermal field theories}
\author{Wenyang Qian}
\emailAdd{qian.wenyang@usc.es}
\author{and Bin Wu}
\emailAdd{b.wu@cern.ch}
\affiliation{Instituto Galego de F\'isica de Altas Enerx\'ias IGFAE, Universidade de Santiago de Compostela,
E-15782 Galicia-Spain}

\abstract{
Thermal properties of quantum fields at finite temperature are crucial to understanding strongly interacting matter and recent development in quantum computing has provided an alternative and promising avenue of study. In this work, we study thermal field theories involving only fermions using quantum algorithms. We first delve into the presentations of fermion fields via qubits on digital quantum computers alongside the quantum algorithms such as quantum imaginary time evolutions employed to evaluate thermal properties of generic quantum field theories. Specifically, we show numerical results such as the thermal distribution and the energy density of thermal field theories for Majorana fermions in 1+1 dimensions using quantum simulators. In addition to free field theory, we also study the effects of interactions resulting from coupling with a spatially homogeneous Majorana field. In both cases, we show analytically that thermal properties of the system can be described using phase-space distributions, and the quantum simulation results agree with analytical and semiclassical expectations. Our work is an important step to understand thermal fixed points, preparing for quantum simulation of thermalization in real time.
}

\maketitle

\section{Introduction}
Quantum information science has proved useful in a broad range of physics applications ranging from simulation of quantum field theories (QFTs)~\cite{Jordan:2011ci, Jordan:2012xnu, Jordan:2014tma, Klco:2018zqz, Kharzeev:2020kgc} and open quantum system~\cite{Klco:2018kyo, deJong:2021wsd} and thermalization~\cite{Zhou:2021kdl, Xie:2022jgj, Du:2023ewh} to nuclear structures~\cite{Kreshchuk:2020dla, Qian:2021jxp, Kreshchuk:2023btr} and hard probes~\cite{Wei:2019rqy, Barata:2023clv, deLejarza:2022bwc} and collider physics~\cite{Gustafson:2022dsq, Bepari:2021kwv, Alvi:2022fkk}, especially thanks to the fast development of quantum computing technology in recent years. Though most of these works are involved with either a small amount of qubits or classical simulation techniques, they have demonstrated that full quantum simulation is potentially more advantageous than classical methods in reducing the problem complexity from exponential to polynomial coupled to large-scale simulation, as expected based on general arguments due to local interactions~\cite{Lloyd:1996aai} and multi-dimensional Hilbert space. Theoretical analysis of quantum simulation algorithm employed in QFTs has already provided such a promise in the next few decades when fault-tolerant quantum computing is available~\cite{Jordan:2012xnu,Jordan:2014tma}. Among these applications, the study of the real-time dynamics and thermalization in quantum field theory on quantum computers or annealers can be potentially of great significance as traditional Monte Carlo sampling techniques are limited by the sign problem~\cite{deForcrand:2009zkb}. 
By trading quantum statistics for classical sampling, quantum simulation provides a natural way to simulate nature~\cite{Feynman:1981tf}.

In the study of thermalization, one is interested to investigate whether and how a quantum field approaches to thermal states over time. This paper focuses on using quantum algorithms to prepare thermal states and subsequently evaluate physical observables in fermionic thermal field theories. Thermal states are mixed states, which are typically difficult to prepare on a circuit and involve non-unitary operations. Recently, novel approaches have been proposed to simulate thermal states on a circuit, including the variational quantum algorithm~\cite{mcardle2019variational}, the quantum imaginary time evolution (QITE)~\cite{Motta:2019yya}, the thermal pure quantum state~\cite{Sugiura:2013pla, Davoudi:2022uzo}, the open-system evolution algorithm~\cite{Cleve:2016dgx}, the thermofield double state approach~\cite{Cottrell:2018ash, Failde:2023bho}, and so forth. Among these approaches, the QITE algorithm, is particularly appealing, for it requires exponentially less space and time per iteration compared with their classical counterparts owing to the information saturation for large systems~\cite{Motta:2019yya}. In addition, unlike many other quantum approaches, it is practical to implement the QITE on current noisy intermediate devices without requiring deep circuits, ancilla qubits, or high-dimensional optimization. Consequently, due to these benefits, the QITE algorithm has been successfully demonstrated to study phase diagrams of strong interactions at finite temperature and chemical potentials~\cite{Czajka:2021yll, Czajka:2022plx, Davoudi:2022uzo}.

In this work, we study fermion fields at finite temperature and density with the QITE algorithm evaluated on a quantum simulator. Unlike previous studies as reviewed above, we focus on evaluating the phase-space distribution in systems amenable to quasiparticle descriptions using quantum algorithms and checking the validity of such a quantum computation by comparing it with analytical results. While our approach to quantum computation is broadly applicable to generic fermionic field theories, this particular investigation allows us to compare with the Fermi-Dirac distribution, which serves as the thermal fixed point for the semiclassical transport description of quarks in quantum chromodynamics (QCD) (see refs.~\cite{Kurkela:2018oqw, Kurkela:2018xxd, Du:2020zqg, Du:2020dvp, Cabodevila:2023htm} for recent studies). Moreover, according to the non-cloning theorem~\cite{Wootters:1982zz}, one has to start over to prepare thermal states after each measurement to compute any new observables.  One may, on the other hand, only measure the phase-space distributions in such systems, which serve as a semiclassical copy of the thermal quantum states. 

In more detail, we demonstrate quantum computation in thermal field theory using Majorana fields~\cite{Elliott:2014iha} in 1+1 dimensions to minimize resources required for quantum simulations. In particular, we study the thermal distribution and the energy density in both the high-temperature and low-temperature limits at thermal equilibrium. In free field theory, which contains only one type of particles, we find the thermal distribution nicely follows the Fermi-Dirac distribution as expected at 4 qubits, but more qubits for the energy density are needed to compare with the continuum. In addition, we study the fermion-fermion interaction with an additional spatially homogeneous Majorana field, which generates an additional type of quasiparticles, to reveal the effects of the interaction. All our simulations are in agreement with exact diagonalization, which suggests quantum computing is a promising tool for studying thermal field theory. 

This manuscript is organized as follows. In sec.~\ref{sec:theory}, we review fermion fields in 1+1 dimensions and establish their qubit encoding schemes on the lattice. We introduce the QITE algorithm used to calculate thermal observables in this work. In sec.~\ref{sec:free_fermion}, we present our simulation results of the thermal distribution and the energy density for the free fermion fields and compare our results with the continuum. In sec.~\ref{sec:interaction_fermion}, we present our simulation results of the fermion fields coupled with another homogeneous background field. In sec.~\ref{sec:summary}, we summarize and discuss future avenues of this work.

\section{Quantum algorithm for fermionic thermal field theories}\label{sec:theory}
In this section, aimed at a broader audience, we make a self-contained, detailed introduction to the study of a quantum field theory (QFT) involving only fermions by approximating it with a discrete many-body system~\cite{Jordan:2014tma}. We focus on the QFT at finite temperature, and our methodology presented below is general, similar to that employed for Dirac fermions in refs~\cite{Czajka:2021yll, Czajka:2022plx, Pedersen:2023asd}. For the sake of concise notation and due to limitations in quantum simulation resources, we instead study Majorana fermions~\cite{Elliott:2014iha} in 1+1 dimensions as an illustration of our approach.

\subsection{Fermion fields in 1+1 dimensions}

In this work, we explicitly study a QFT for
Majorana fermions in 1+1 dimensions with the Lagrangian density given by
\begin{align}\label{eq:lagrangian}
    \mathcal{L} = \frac{1}{2}\bar{\psi}(i\slashed{\partial}-m)\psi - \mathcal{H}_I(\psi),
\end{align}
where $\mathcal{H}_I$ is the interaction Hamiltonian, and $\gamma^{\mu}$ is chosen to be in the Majorana representation~\cite{Peskin:1995ev}:
\begin{align}
    \gamma^0=\begin{pmatrix}0 & -i\\i &0\end{pmatrix},\qquad \gamma^1 = \begin{pmatrix}
    0 & i\\
    i & 0
    \end{pmatrix}.
\end{align}
Here, the Majorana field $\psi$ satisfies $\psi^\dagger = \psi^T$. The field theory can be quantized by imposing the canonical anticommutation relations
\begin{align}
    \{\psi^{\alpha}(t, x), \psi^{\beta}(t,y)\}=\delta(x-y)\delta^{\alpha\beta}.
\end{align}
In the following discussion, $\mathcal{H}_I$ is considered to be generic and is assumed to be independent of the time derivative of the Majorana field.

From the Lagrangian density in eq.~\eqref{eq:lagrangian}, the energy-momentum tensor of the theory is given by
\begin{align}
    \hat{T}^{\mu\nu} = \frac{1}{2} \bar{\psi}i\gamma^{\mu}\partial^{\nu}\psi - g^{\mu\nu}\mathcal{L}
\end{align}
with the metric tensor $g^{\mu\nu} = g_{\mu\nu} =\mathrm{diag}(1, -1)$. Its components correspond respectively to the energy density $\hat{\epsilon}$, the pressure $\hat{P}$, and the momentum operator $\hat{p}$:
\begin{align}\label{eq:TComp}
    \hat{\epsilon}\equiv T^{00},\qquad \hat{P}\equiv T^{11}, \qquad\hat{p} = - \frac{i}{2}\int dx \bar{\psi}\gamma^1\partial_x\psi.
\end{align}
And the Hamiltonian is given by
\begin{align}\label{eq:hamiltonian_from_lagrangian}
    \hat{H} = \int dx \epsilon =  \frac{1}{2}\int dx[\bar{\psi}(-i\gamma^1\nabla_{\mathbf{x}}+m)\psi] + \hat{H}_I
\end{align}
with $\hat{H}_I\equiv\int dx \mathcal{H}_I(\psi)$.

In the Heisenberg picture, the time evolution of the field operator is governed by the Heisenberg equation of motion:
\begin{align}\label{eq:timeEvol}
    \dot{\psi}(t,x)=i[\hat{H},\psi(t,x)],
\end{align}
where the overdot denotes the derivatives with respect to time $t$. In free field theory, the solution to the above equation can be written in the form
\begin{align}
    \psi(t, x) = \int\frac{d{{p}}}{2\pi}\frac{1}{\sqrt{2E_{{p}}}}(\hat{a}_p u_p e^{-i E_p t + ipx}+\hat{a}_p^\dagger v_p e^{iE_p t - ipx}),
\end{align}
where the spinors in the Majorana representation are given by
\begin{align}\label{eq:spinors}
    u_p = \left(\begin{array}{c}
         \sqrt{p^+}  \\
         i\sqrt{p^-}
    \end{array}\right),\qquad
    v_p= u^*_p=\left(\begin{array}{c}
         \sqrt{p^+}  \\
         -i\sqrt{p^-}
    \end{array}\right)    
\end{align}
with $p^\pm\equiv E_{{p}}\pm{p}$, and the creation and annihilation operators satisfy
\begin{align}
    \{\hat{a}_{{p}}, \hat{a}^\dagger_{{p'}}\}=2\pi\delta({p}-{p}'),\qquad \{\hat{a}_{{p}}, \hat{a}_{{p'}}\}=\{\hat{a}^\dagger_{{p}}, \hat{a}^\dagger_{{p'}}\}=0.
\end{align}
Note that all the discussions here and below can be straightforwardly generalized into cases involving Dirac fields in four dimensions. In such cases, the creation operator $\hat{a}^\dagger_p$ is replaced by the creation operator for the antifermion in three spatial dimensions~\cite{Peskin:1995ev}.

\subsection{Representation of fermion fields using qubits}
\label{sec:discretization}
In this subsection, following ref.~\cite{Jordan:2014tma}, we approximate the continuum fermion field theory in eq.~(\ref{eq:lagrangian}) by a discrete theory on a spatial lattice, and present the fermion fields using qubits. 

\subsubsection{Fermion fields on a spatial lattice}
\label{sec:RepCoordinate}
In this paper, the continuum field theory is approximated by a discrete theory similar to that in lattice QCD (see, e.g., ref.~\cite{DeGrand:2006zz}), with the only difference being that time remains continuous. The fermion field $\psi_n(t)\equiv \sqrt{a}\psi(t, na)$ is now assumed to live on a spatial lattice with
\begin{align}\label{eq:lattice_discretization_x}
    x \in \Omega_x \equiv\{ 0, a, \cdots, (N-1)a\},
\end{align}
satisfying periodic boundary conditions $\psi_N(t)=\psi_0(t)$. Here, $a$ is the lattice spacing and $N$ is a positive integer. Accordingly, the Fourier transform of any function $f$ in the continuum theory is replaced by its discrete Fourier transform, and the momentum $p$ is cut off at both the infrared (IR) and the ultraviolet (UV) limits. To be more specific, the IR cutoff of $p$ follows from the fact that all the quantities in this theory are the same at $x$ and $x+L$ with $L\equiv aN$, yielding $p L = 2\pi n$ with $n$ an integer. That is, the softest modes (besides the zero mode) are given by $|p|=\Delta p\equiv 2\pi/L$. The UV cutoff results from the lattice spacing, yielding only $N$ distinct values of $p$, which can be chosen to be as follows\footnote{
Translations to other intervals such as $ \Omega_p = \{0, \Delta p, \cdots, (N - 1)\Delta p\}$ make no difference in the definition of the discrete Fourier transform but not all the values in this interval corresponds to physical momenta in the limit $a\to 0$. 
}
\begin{align}\label{eq:lattice_discretization}
    \Omega_p \equiv\left\{\begin{array}{ll}
         \{(\frac{-N+1}{2})\Delta p, \cdots, (\frac{N-1}{2})\Delta p\} &\text{if $N$ is odd} \\
         \{(\frac{-N+2}{2})\Delta p, \cdots, (\frac{N}{2})\Delta p\}&\text{if $N$ is even} 
    \end{array}\right.
\end{align}
That is, the UV cutoff of $|p|$ is around $\pi/a$.

The discrete theory is not unique and we choose to construct the discrete theory by making the following replacement for any function $f$ and $x=na$: 
\begin{align}\label{eq:replacement}
    &\int dx f(x) \to a \sum_{n=0}^{N-1}f(na),\notag\\
    \partial_x \psi(x)\to \frac{1}{\sqrt{a}}&\partial_x \psi_n\equiv \frac{1}{\sqrt{a}} \frac{\psi_{n+1}-\psi_{n-1}}{2a}, \quad \psi(x)=\sqrt{a}\psi_n
\end{align}
in the continuum Lagrangian. In this case, one obtains the Lagrangian for the discrete theory as follows
\begin{align}\label{eq:L_Dis}
          L = \sum_{n=0}^{N-1}[\frac{1}{2}\bar{\psi}_n(i\slashed{\partial}-m)\psi_n - a\mathcal{H}_I(\psi_n/\sqrt{a})]\equiv \frac{1}{2}\sum_{n=0}^{N-1}\bar{\psi}_n(i\slashed{\partial}-m)\psi_n - \hat{H}_I.
\end{align}
After the canonical quantization of $L$, all the dynamics in the discrete quantum theory can be in principle derived. The continuum theory is expected to be recovered in the limit $a\to 0$ and $L=aN\to\infty$, which is a non-trivial limit for interacting theories that involves the implementation of renormalization.

In detail, the fermion field is quantized on the lattice, satisfying
\begin{align}\label{eq:anticommutingRelation}
    \{\psi^{\alpha}_{n}, \psi^{\beta}_{n'}\}=\delta^{\alpha\beta}\delta_{nn'},
\end{align}
where $\delta_{nn'}=1$ only if $n-n'=0~\text{Mod}~N$. The Hamiltonian corresponding to $L$ in eq.~\eqref{eq:L_Dis} takes the form
\begin{align}\label{eq:H_with_pos_fields}
    \hat{H} =&\frac{1}{2}\sum\limits_n \bar{\psi}_{n}[-\frac{i}{2a}\gamma^1(\psi_{n+1}-\psi_{n-1})+m\psi_{n}]\notag\\ &-\frac{r}{4a}\sum\limits_n\bar{\psi}_n(\psi_{n+1} - 2\psi_n + \psi_{n-1}) + \hat{H_I},
\end{align}
where the so-called Wilson term with $0< r\leq 1$ is included to prevent fermion doubling~\cite{Wilson:1974sk}.\footnote{Alternatively, one may use staggered fermions~\cite{Kogut:1974ag, Susskind:1976jm}.} In the Heisenberg picture, the field operators $\psi_n$ at any time $t$ satisfy the Heisenberg evolution equation of motion, that is, eq.~(\ref{eq:timeEvol}) with the Hamiltonian replaced by the above discrete Hamiltonian. To study the energy density and the pressure in thermal systems on the discretized space, one should consistently rewrite the energy-momentum tensor $\hat{T}_n^{\mu\nu}$ in eq.~(\ref{eq:TComp}) using replacement in eq.~(\ref{eq:replacement}). Additionally, the time derivative of $\psi_n$ in the expression of $\hat{T}_n^{\mu\nu}$ is eliminated by employing the Heisenberg evolution equation of motion.

In this work, we use the quantum circuit model for quantum computations~\cite{Nielsen:2012yss} to carry out simulations. Unlike the Dirac fermions, as discussed in refs.~\cite{Jordan:2014tma, Czajka:2021yll, Xie:2022jgj, Pedersen:2023asd}, we need only $N$ qubits here to present $N$ Majorana fields either in coordinate space or momentum space. As we are dealing with fermion fields, it is convenient to construct a complete set of commuting fermionic number operators, and assign the two eigenstates of each operator to a single qubit.

\subsubsection{Representation in coordinate space}
The anticommutation relations for the Majorana fermions in eq.~(\ref{eq:anticommutingRelation}) are different from those for the Dirac fermions. In order to construct the complete set of the fermionic number operators, we define
\begin{align}
\psi_n =\frac{1}{\sqrt{2}}\begin{pmatrix}
    1&1\\
    i&-i
\end{pmatrix}
\begin{pmatrix} a_n\\ a^\dagger_n \end{pmatrix}
,
\end{align}
where the new fields obey the following anticommutation relations
\begin{align}\label{eq:anticommutingRelationFora}
    \{a_n, a^\dagger_{n'}\}=\delta_{nn'},\qquad \{a_n, a_{n'}\}=0,\qquad \{a^\dagger_n, a^\dagger_{n'}\}=0.
\end{align}
In terms of these new fields, the Hamiltonian takes the form
\begin{align}\label{eq:HMajorana}
    \hat{H} =\hat{H}_0+\hat{H}_I
\end{align}
with the free Hamiltonian given by
\begin{align}
    \hat{H}_0=&\sum\limits_n\bigg[-i \frac{a_n a_{n+1} + a^\dagger_n a^\dagger_{n+1}}{2a} + m\bigg( a_n^\dagger a_n - \frac{1}{2}\bigg)\bigg]\notag\\
    &- \frac{r}{2a} \sum\limits_n[a_n^\dagger(a_{n+1} - 2 a_n + a_{n-1}) +1].
\end{align}
That is, the continuum quantum field described in eq.~(\ref{eq:lagrangian}) is now approximated by a many-body system consisting of $N$ fermions, whose behavior is governed by the above Hamiltonian.

 In coordinate space, we choose the complete set of commuting fermionic number operators as $\{a_n^\dagger a_n|n=0,\cdots, N-1\}$ and the corresponding complete orthonormal basis of the Hilbert space is given by
\begin{align}
    &|n_0 n_1\cdots n_{N-1} \rangle_x=\prod_{i=0}^{N-1}(a_i^\dagger)^{n_i}|0\rangle_x\equiv (a^\dagger_0)^{n_0}(a^\dagger_1)^{n_1}\cdots(a^\dagger_{N-1})^{n_{N-1}}|0\rangle_x,\notag\\
    &{}_x\langle n_0 n_1\cdots n_{N-1}|={}_x\langle 0 |\prod^{0}_{i = N-1}(a_i)^{n_i}\equiv {}_x\langle 0 |(a_{N-1})^{n_{N-1}}(a_{N-2})^{n_{N-2}}\cdots(a_0)^{n_0},
\end{align}
where $n_i=0, 1$ and the vacuum state in coordinate space $|0\rangle_x$ is defined by $a_n|0\rangle_x=0$ for all $n\in\Omega_x$. Note that we add a subscript $x$ to the state vectors as a reminder that this vacuum state is not necessarily the same as the physical vacuum state which is defined as the ground state of $\hat{H}$. One can then map the $N$ fermions to $N$ qubits using either the Jordan-Wigner~\cite{Jordan:1928wi} or the Bravyi-Kitaev~\cite{bravyi2002fermionic} transformations. For simplicity, we use the Jordan-Wigner transformation to represent the two states $(a^\dagger_i)^{n_i}|0\rangle$ with $n_i=0, 1$ using the $i^{\text{th}}$ qubit. In this case we have
\begin{align}\label{eq:jordanwigner}
    a_n^\dagger = \frac{\sigma^X_n - i \sigma^Y_n}{2} \prod_{i=0}^{n-1}\sigma^Z_i,\quad\quad
    a_n = \frac{\sigma^X_n + i \sigma^Y_n}{2} \prod_{i=0}^{n-1}\sigma^Z_i,
\end{align}
where the $\sigma^X_n, \sigma^Y_n, \sigma^Z_n$ are the $2\times 2$ Pauli matrices acting on the $n^{\text{th}}$ qubit. 

In this paper, we focus on properties of spatially homogeneous systems in thermal equilibrium. As we shall show in subsequent sections, the periodical boundary conditions imposed on the fermion fields with $\psi_{N} = \psi_0$ are well suited for such studies. This is a consequence of the fact that the Hamiltonian commutes with any spatial translation, that is, a displacement of the fields by any spatial off-set: $a \Delta n$ with $\Delta n$ an integer. This symmetry corresponds to the momentum conservation in the continuum limit. Let us define the translation transformation as
\begin{align}
    U^\dagger_m a_n U_m = a_{n+m},\qquad
    U^{\dagger}_m a^{\dagger}_n U_m = a^\dagger_{n+m},\qquad U^\dagger_m \psi_n U_m=\psi_{n+m}.
\end{align}
Evidently, one has $U^\dagger_m \hat{H} U_m = \hat{H}$. The translation transformations are unitary with $U_m^{-1} = U_m^\dagger = U_{-m}$. This can be easily checked as follows. Given
\begin{align}
    U_m^\dagger|0\rangle = |0\rangle,
\end{align}
one has
\begin{align}
    U_m^{\dagger}|n_0, n_1,\cdots, n_{N-1} \rangle = (-1)^{n_{perm}}|n_{N-m},\cdots, n_{N-1}, n_0, n_1,\cdots, n_{N-m-1} \rangle,
\end{align}
where the number of permutations of the creation operators is given by
\begin{align}
    n_{perm}= \bigg( \sum\limits_{i=0}^{N-m-1} n_{i} \bigg) \bigg(\sum\limits_{j = N-m}^{N-1} n_{j}\bigg).
\end{align}
That is, it maps one basis state vector into another up to a phase and is therefore unitary.

\subsubsection{Representation in momentum space}
\label{sec:RepMomentum}
In momentum space, we choose the complete orthonormal basis
of the Hilbert space as the eigenstates of the momentum operator corresponding to the spatial translation invariance.
By substituting eq.~(\ref{eq:replacement}) into the definition of $\hat{p}$ for the continuum theory, as given by eq.~(\ref{eq:TComp}), we define the momentum operator for the discrete theory as
\begin{align}
    \hat{p}\equiv - \frac{i}{4a}\sum_{n}\bar{\psi}_n\gamma^1(\psi_{n+1} - \psi_{n-1}) = -\frac{i}{2a}\sum\limits_n a_n^{\dagger} (a_{n+1} - a_{n-1}).
\end{align}
Using the anticommutation relations as given in eq.~(\ref{eq:anticommutingRelationFora}), one has
\begin{align}
    [\hat{p}, a_n] = \frac{i}{2a}(a_{n+1} - a_{n-1}),\qquad [\hat{p}, a^\dagger_n] = \frac{i}{2a}(a^\dagger_{n+1} - a^\dagger_{n-1}),
\end{align}
and, consequently,\footnote{
Here, we assumes $[\hat{p},\hat{H}_I]=0$, which is the case for the interaction Hamiltonian chosen in sec.~\ref{sec:interaction_fermion}.
}
\begin{align}
    [\hat{p},\hat{H}]=0.
\end{align}
Therefore, one may use the eigenstate of $\hat{p}$ to distinguish the degenerate energy eigenstates. 

In terms of the momentum operator $\hat{p}$, the momentum eigenstates can be constructed by using
\begin{align}
    \Tilde{a}^\dagger_{p}\equiv \frac{1}{\sqrt{N}} \sum\limits_n e^{i p a n} a_n^\dagger,\qquad
    \Tilde{a}_{p}\equiv \frac{1}{\sqrt{N}} \sum\limits_n e^{-i p a n} a_n,
\end{align}
which satisfy the following anticommutation relations
\begin{align}
    \{\tilde{a}_p, \tilde{a}^\dagger_{p'}\}=\delta_{pp'},
    \qquad
    \{\tilde{a}_p, \tilde{a}_{p'}\}=0,
    \qquad
    \{\tilde{a}^\dagger_p, \tilde{a}^\dagger_{p'}\}=0.
\end{align}
Accordingly, one has
\begin{align}
    [\hat{p},\tilde{a}^{\dagger}_{p}] = \frac{\sin(pa)}{a}\tilde{a}^{\dagger}_{p},\qquad [\hat{p},\tilde{a}_{p}] = - \frac{\sin(pa)}{a}\tilde{a}_{p},
\end{align}
and
\begin{align}
    U_m^\dagger\tilde{a}^{\dagger}_{p} U_m = \tilde{a}^{\dagger}_{p} e^{-ipam},\qquad 
    U_m^\dagger\tilde{a}_{p} U_m = \tilde{a}_{p} e^{ipam}.
\end{align}
As a result, the eigenstate of $\hat{p}$ corresponding to the eigenvalue $\sin(pa)/a$, denoted as $|p\rangle$, can be chosen as any superposition of 
\begin{align}
    \tilde{a}^{\dagger}_{p} |0\rangle\qquad\text{and}\qquad \tilde{a}_{-p} |0\rangle
\end{align}
with
\begin{align}
    \hat{p}|p\rangle = \frac{\sin(pa)}{a}|p\rangle,
    \qquad
    U_m |p\rangle = e^{i p a m}|p\rangle.
\end{align}

In order to diagonalize the free Hamiltonian, we construct the creation and annihilation operators in momentum space in terms of $\tilde{a}_p^\dagger$ and $\tilde{a}_p$ through the following transformations 
\begin{align}\label{eq:ahat_p}
    \begin{pmatrix}       \hat{a}^\dagger_p\\\hat{a}_{-p}
    \end{pmatrix}
    =\frac{1}{2\sqrt{E_p}}
    \begin{pmatrix}
       \sqrt{p^+} + \sqrt{p^-} & \sqrt{p^+} - \sqrt{p^-}\\
       \sqrt{p^-} - \sqrt{p^+} & \sqrt{p^+} + \sqrt{p^-}       
    \end{pmatrix}
    \begin{pmatrix}       \tilde{a}^\dagger_p\\\tilde{a}_{-p}
    \end{pmatrix}    
\end{align}
with
\begin{align}\label{eq:Epas}
    E_{p}=\sqrt{\bigg(m+\frac{2r}{a}\sin^2\frac{a{p}}{2}\bigg)^2 + p_a^2}, \qquad    {p}_a=\frac{1}{a}\sin({p}a),
    \qquad
    p^\pm \equiv E_p \pm p_a.
\end{align}
It is easy to check that these creation and annihilation operators satisfy the following anticommutation relations:
\begin{align}\label{eq:anticommutingRelationForap}
    \{\hat{a}_p, \hat{a}^\dagger_{p'}\}=\delta_{pp'},
    \qquad
    \{\hat{a}_p, \hat{a}_{p'}\}=0,
    \qquad
    \{\hat{a}^\dagger_p, \hat{a}^\dagger_{p'}\}=0.
\end{align}
Accordingly, we choose the complete set of commuting fermionic number operators as $\{\hat{a}_p^\dagger \hat{a}_p|p\in\Omega_p\}$, and the complete orthonormal basis
\begin{align}
    |\{n_p p|p\in\Omega_p\}\rangle_p \equiv \prod_{p\in\Omega_p}(\hat{a}_p^\dagger)^{n_p}|0\rangle_p ,
\end{align}
where $n_p=0$ or $1$, and the vacuum state in momentum space is defined by $\hat{a}_p|0\rangle_p=0$ for all $p\in \Omega_p$.

By inverting eq.~(\ref{eq:ahat_p}), one has 
\begin{align}
    \begin{pmatrix}       \tilde{a}^\dagger_p\\\tilde{a}_{-p}
    \end{pmatrix}
    =\frac{1}{2\sqrt{E_p}}
    \begin{pmatrix}
       \sqrt{p^+} + \sqrt{p^-} & \sqrt{p^-} - \sqrt{p^+}\\
       \sqrt{p^+} - \sqrt{p^-} & \sqrt{p^+} + \sqrt{p^-}       
    \end{pmatrix}
    \begin{pmatrix}       \hat{a}^\dagger_p\\\hat{a}_{-p}
    \end{pmatrix}.
\end{align}
Accordingly, the fermion field takes the compact form
\begin{align}\label{eq:Psin}
    \psi_n & \equiv \sum\limits_p
    \frac{1}{\sqrt{2 N E_{p}}}[\hat{a}_p u_p e^{ipan} + \hat{a}_p^\dagger v_p e^{-ipan}]
\end{align}
and, conversely, one has
\begin{align}\label{eq:aAndPsi}
     \hat{a}_p = \frac{1}{\sqrt{2N E_p}} \sum\limits_n e^{-ipan} \bar{u}_p\gamma^0\psi_n,
     \qquad
     \hat{a}^\dagger_p = \frac{1}{\sqrt{2N E_p}} \sum\limits_n e^{ipan} \bar\psi_n\gamma^0 u_p
\end{align}
with the energy and the momentum given in eq.~(\ref{eq:Epas}). One can easily check that the free Hamiltonian takes the form
\begin{align}\label{eq:H0}
    \hat{H}_0\equiv \sum\limits_p E_p (\hat{a}^\dagger_p \hat{a}_p - 1/2).
\end{align}
Note that the above derivation is valid irrespective of the form of $\hat{H}_I$. And $|p\rangle$ is the momentum eigenstate even for interacting theories, though $|p\rangle$ may not correspond to one single-particle state in general. 

\subsection{Fermionic thermal fields}

Using the discrete theory introduced in the previous subsection, one can straightforwardly study properties of the $N$-fermion systems in thermal equilibrium, as an approximation to the underlying thermal field theory~\cite{Laine:2016hma}. In terms of the partition function
\begin{align}\label{eq:part_func}
    Z_\beta \equiv \text{Tr}[e^{-\beta \hat{H}}],
\end{align}
where $\beta=1/T$ with $T$ the temperature, the expectation value of any observable $\hat{O}$ is given by
\begin{align}\label{eq:thermal_avg}
    \langle\hat{O}\rangle_\beta\equiv Z_\beta^{-1}{\text{Tr}[e^{-\beta  \hat{H}} \hat{O}]}.
\end{align}
Here, the trace can be carried out by summing the expectation values over the complete orthonormal states either in coordinate space or in momentum space. 

The trace performed in terms of the orthonormal bases in the previous section is designed to study spatially homogeneous systems. Accordingly, the expectation value of any local operator $\hat{O}_n$ should be independent of the spatial position $n$. This can be easily checked by employing the translation invariance of $\hat{H}$, i.e., $[U_m,\hat{H}]=0$:
\begin{align}\label{eq:spatialHomo}
    \langle \hat{O}_n\rangle_\beta=Z_\beta^{-1} \text{Tr}[e^{-\beta \hat{H} }  U^\dagger_n \hat{O}_0 U_n]=Z_\beta^{-1}\text{Tr}[U^\dagger_n e^{-\beta \hat{H} } 
    \hat{O}_0 U_n]=Z_\beta^{-1}\text{Tr}[e^{-\beta \hat{H} } \hat{O}_0].
\end{align}
In this work, we focus on the energy density $\hat{\epsilon}_n\equiv \hat{T}_n^{00}$.

Similarly, for the correlation function of any two local operators, one has
\begin{align}
    \langle \hat{O}_n \hat{O}'_m\rangle_\beta
    =
    \langle \hat{O}_{n-m} \hat{O}'_0\rangle_\beta.
\end{align}
Accordingly, one can Fourier transform it similarly as in the continuum limit. For example, the spectral function is now defined as
\begin{align}\label{eq:spectrualFunc}
    \rho(p^0, p)\equiv \sum\limits_n\int dt e^{ip^0 t - i p a n}\langle\{\psi_n(t), \bar{\psi}_0(0)\} \rangle_\beta
\end{align}
with $p\in \Omega_p$. In the QFTs to be studied in the following sections, one has
\begin{align}
    \rho(p^0, p) = 2\pi\,\text{sign}(p^0)\sum_i P_i \delta((p^0)^2-(E_p(m_i))^2)(\slashed{p}+m_i),
\end{align}
where $m_i$ denotes the mass of the $i^{\text{th}}$ quasiparticle, $P_i$ stands for the probability to excite the $i^{\text{th}}$ quasiparticle, and the dependence of $E_p$ on mass $m_i$, as defined in eq.~(\ref{eq:Epas}), is explicitly indicated as $E_p(m_i)$. In this case, we define the phase space distribution of the $i^{\text{th}}$ quasiparticle as
\begin{align}
    f^i_p & = \langle \hat{a}^{i\dagger}_{p}\hat{a}^i_{p}\rangle_\beta,
\end{align}
where the creation and annihilation operators $\hat{a}^{i\dagger}_{p}$ and $\hat{a}^i_{p}$ are defined through some transformations of $a^\dagger_n$ and $a_n$, the same as those to define $\hat{a}_p^\dagger$ and $\hat{a}_p^\dagger$ in the previous section with $m$ replaced by $m_i$.

\subsection{Quantum algorithm for simulating thermal fields}\label{sec:qite}

In this work, our goal is calculating thermal averages of physical observables following eq.~\eqref{eq:thermal_avg} for the fermion fields at finite temperature. While it is true for small systems one can directly evaluate thermal averages $\braket{\hat{O}}_\beta$ using trace through exact diagonalization of the Hamiltonian, it becomes increasingly expensive, as $\mathcal{O}((2^N)^3)$, for large systems of $N$ qubits with many degrees of freedom. It is often more favorable to reformulate the problem as an path-integral and use Monte Carlo methods with importance sampling for evaluation. However, the sign problem can arise if any subpath $\braket{x_i|e^{-\beta\hat{H}/N}|x_{i+1}}$ in the integral becomes negative or imaginary, such as when dealing with real-time dynamics~\cite{Kharzeev:2020kgc}, non-zero chemical potentials~\cite{Czajka:2021yll, Czajka:2022plx}, and topological terms~\cite{Honda:2021aum}. Quantum simulation techniques, which deal with quantum statistics instead and operate within the multi-dimensional Hilbert space, are widely believed to be compelling solutions for addressing this intrinsic computational challenges without encountering the sign problem. 

However, preparing thermal states on quantum circuits is generally challenging due to the inherent unitary nature of quantum evolution. To achieve non-unitary evolutions for the thermal mixed state, one would use specialized techniques on the circuit, involving a combination of ancilla qubits, partial measurement, and post-processing. Hybrid strategies using variational circuits can be useful to extract thermal properties on near-term devices. The variational quantum thermalizer algorithm~\cite{verdon2019quantum, Selisko:2022wlc, Fromm:2023npm} utilizes multiple variational circuits to minimize the Helmholtz free energy and study Gibbs states. The McLachlan's variational principle~\cite{Yuan2019theoryofvariational, McLachlan1964, Yao:PRX2021, Chen:2023skq} can also facilitate the evolution in imaginary time to construct variational thermal states. Besides variational methods, the thermal states can be prepared using open quantum system approaches~\cite{Cleve:2016dgx} based on the Stinespring dilation theorem, as demonstrated in the studies of the thermalization of the Schwinger model~\cite{deJong:2021wsd,DeJong:2020riy}. Other promising simulation strategies for preparing a mixed thermal state include the quantum active cooling~\cite{Ball:2022dxy, Cohen:2021imf} and thermofield double states~\cite{Cottrell:2018ash, Zhu:2019bri, Wu:2018nrn}. Importantly, successful simulation of mixed states on quantum circuits typically involve an integration of several of these different methods, tailored to the specific problem at hand.

Rather than examining the efficiency of all the quantum algorithms mentioned above, in our practical simulation of a 1+1 dimension fermion system, we employ the quantum imaginary time evolution (QITE) algorithm~\cite{Motta:2019yya}, which we will briefly review below. This serves to demonstrate and verify the thermal state properties that are analytically evaluated in the following sections. Another key reason for choosing this algorithm is its close relation to the real-time dynamics we plan to explore in our future studies.

\subsubsection{Quantum imaginary time evolution}

We use the QITE algorithm~\cite{Motta:2019yya} to prepare the thermal state and the thermal expectation value of an observable $\hat{O}$ at temperature $T=1/\beta$ is evaluated as
\begin{align}\label{eq:qite_observable}
    \braket{\hat{O}}_\beta &=\frac{\mathrm{Tr}\big[e^{-\beta \hat{H}} \hat{O}\big]}{\mathrm{Tr}\big[e^{-\beta \hat{H}}\big]}\;,
\end{align}
with the two traces 
\begin{align}\label{eq:qite_trace}
    \mathrm{Tr}\big[e^{-\beta \hat{H}} \hat{O}\big] =  \sum_{k\in \mathcal{S}}\braket{k|e^{-\beta \hat{H}/2} | \hat{O} | e^{-\beta \hat{H}/2}|k},\quad \mathrm{Tr}\big[e^{-\beta \hat{H}}\big]=
    \sum_{k\in \mathcal{S}}\braket{k|e^{-\beta \hat{H}}|k}
\end{align}
that typically sum over the entire computational basis $\mathcal{S}$\footnote{The naive set of computational basis is the classical product states, i.e., $\{\ket{0}, \ket{1}\}^{\otimes N}$ over $N$ qubits.}.
Given any initial state $k$, the final state $\ket{\psi_k(\beta/2)} = e^{-\beta \hat{H}/2}\ket{k}$ is obtained by dividing $\beta$ into $N_s$ many small steps $\Delta \beta = \beta / N_s$ using the Trotter formula~\cite{Nielsen:2012yss},
\begin{align}
    \ket{\psi_k(\beta/2)}=e^{-\beta \hat{H}/2} \ket{k}= \prod_{i=1}^{N_s}e^{-\Delta \beta \hat{H}/2} \ket{k}.
\end{align}
For an intermediate state $\ket{\psi_{k}(\beta_i/2)}$ with $\beta_i=i\Delta \beta$ at step $i$, we write it as $\ket{\psi_{k}^i}$ for simplicity. To obtain its subsequent state, the non-unitary operation in $\hat{H}$ is substituted with a unitary evolution of a new operator $\hat{A}_k^i$ to obtain the subsequent state
\begin{align}\label{eq:imaginary_time_approximation}
    \ket{\psi_{k}^{i+1}} = e^{-\Delta \beta \hat{H}/2} \ket{\psi_{k}^i} = \sqrt{c_k^i(\Delta \beta)} e^{-i\Delta \beta \hat{A}_k^i/2} \ket{\psi_{k}^i} + \mathcal{O}(\Delta \beta^2),
\end{align}
with the normalization $c_k^i(\Delta \beta)=\braket{\psi_k^i | e^{-\Delta \beta \hat{H}} | \psi_k^i } \approx 1 - \Delta \beta \braket{\psi_k^i|\hat{H}|\psi_k^i}$. The real-time Hamiltonian operator, $\hat{A}_k^i = \sum_I a_I \sigma_I$ with $\sigma_I$ in the generalized $N$-qubit Pauli bases $\mathcal{B}^{N} = \{I, \sigma^X,\sigma^Y,\sigma^Z\}^{\otimes{N}}$, is determined by solving the linear matrix equation
\begin{align}
    (\bm{S} + \bm{S}^T) \bm{a}_i &= \bm{b}\;,
\end{align}
with matrix elements $b_I = -i \braket{\psi_k^i| \big(\hat{H} \sigma_I - \sigma_I^\dagger \hat{H} \big)|\psi_k^i}/\sqrt{c_k^i(\Delta \beta)}$ and $S_{IJ} = \braket{\psi_k^i | \sigma^\dagger_I \sigma_J|\psi_k^i}$ evaluated on the quantum circuit as expectations. Notably, since the determination of $a_I$ for the real-time evolution operator is systematic and exact, the QITE algorithm does not suffer optimization issues that are typically presented in variational approaches. 

\begin{figure}
    \centering
    \subfigure[\label{fig:QMETTS} QMETTS algorithm
    ]{
    \includegraphics[width=0.44\textwidth]{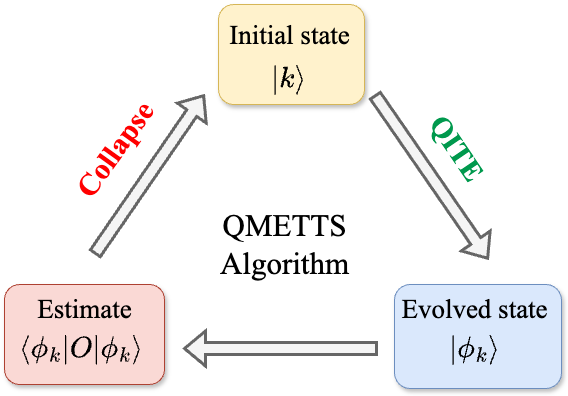}
    }
    \subfigure[\label{fig:QITE} QITE algorithm
    ]{
    \raisebox{2em}{
        \includegraphics[width=0.50\textwidth]{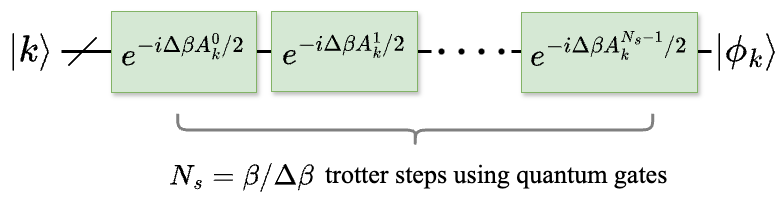}
    }
    }
    \caption{Skematic diagram of the workflow for the QMETTS and QITE algorithms~\cite{Motta:2019yya} used in this work. Note the QITE is a subroutine in the QMETTS. 
    }
    \label{fig:qmetts}
\end{figure}

For $N$-qubit local Hamiltonians, such as the fermionic Hamiltonian we study, the QITE algorithm is especially advantageous as the domain size of the unitary transformation are bounded by $\mathcal{O}(C)$ when assuming a finite correlation length of $C$ qubits for the quantum state. For large systems, this correlation length $C$ generally increases with trotterization and eventually saturates to $C_\mathrm{max} \ll N$~\cite{Hastings:2005pr}. It means that we can potentially work with a much smaller subspace to evaluate each trotter step and obtain the final state, making the quantum algorithm exponentially efficient compared to its classical counterpart~\cite{Motta:2019yya}. Nonetheless, in the small test case studied in this paper, the advantage is not obvious and we will leave it for a future study on larger fermionic systems. Several improvements can be made, such as dividing the expectation evaluations into groups~\cite{Pedersen:2023asd}, trotterizing relevant Hamiltonian terms separately, and using random states to minimize the total number of expectation evaluations~\cite{Davoudi:2022uzo}.

\subsubsection{Thermal expectation on the circuit}

Thermal properties of the fermion fields at finite temperature can be calculated using the QITE algorithm in eq.~\eqref{eq:qite_observable} by post-processing the respective expectation value from each initial state $\ket{k}$. 
Additionally, the QITE evolution can be incorporated into the minimally entangled typical thermal state method (METTS)~\cite{METTS, stoudenmire2010minimally} to sample the thermal observable efficiently. Essentially, we redefine the quantum state as
\begin{align}
\ket{\phi_k}=P_k^{-1/2}e^{-\beta \hat{H}/2} \ket{k}, \quad P_k=\braket{k | e^{-\beta \hat{H}} |k}.
\end{align}
In the QITE with $N_s$ trotterization steps, they simply are
\begin{align}
    \ket{\phi_k} \equiv \prod_{i=1}^{N_s} e^{-i\Delta \beta \hat{A}_k^i/2}\ket{k},\qquad
    P_k \equiv \prod_{i=1}^{N_s} c_k^i(\Delta \beta).
\end{align}
Collectively, the thermal observable in eq.~\eqref{eq:qite_observable} becomes
\begin{align}
    \braket{\hat{O}}_\beta = \sum_{k\in \mathcal{S}} \frac{P_k}{Z} \braket{\hat{O}_k}_\beta,\qquad \braket{\hat{O}_k}_\beta=\braket{\phi_k|\hat{O}|\phi_k},\qquad Z=\sum_{k\in \mathcal{S}} P_k,
\end{align}
which is equivalent as sampling $\ket{\phi_k}$ with probability $P_k/Z$ and summing its expectation $\braket{\hat{O}_k}_\beta$. Since state collapsing (i.e. measurement) automatically 
yields a transition to a new state $k'$ with probability $\mathrm{Prob}(k\rightarrow k') = |\braket{k' | \phi_k}|^2\sim P_{k'}/Z$, we can approximate the thermal average with $N_S \ll |\mathcal{S}|$ samples
\begin{align}
    \braket{\hat{O}}_\beta \approx \frac{1}{N_S}\sum_{k=1}^{N_S} \braket{\hat{O}_k}_\beta
\end{align}
since each measurement favors the most important contributions for the targeted observable. The complete workflow (see fig.~\ref{fig:qmetts}) for the quantum METTS algorithm (QMETTS) is summarized as the following:
\begin{enumerate}
    \item Choose a classical product state $\ket{k}$.
    \item Evolve with the QITE to compute $\ket{\phi_k}$ and evaluate the expectation $\braket{\hat{O}_k}_\beta$.
    \item The state $\ket{\phi_k}$ collapses to a new state $\ket{k'}$ once measured and repeat step 2.
\end{enumerate}
In practice, one may also collapse the state to X-basis and Z-basis in alternating steps to improve numerical stability. In our calculation with low number of qubits, the strength of the QMETTS algorithm is not obvious and we simply evaluate over all possible $\ket{k}\in \mathcal{S}$ in our numerical simulation. Besides using QMETTS, one may also consider adiabatic state preparation protocol~\cite{Hejazi:2023fiq} to competitively improve the convergence rate. Furthermore, the QITE algorithm has also been incorporated into simulating the open quantum system via the Lindblad equation~\cite{Kamakari:2021scj}. These new efforts will further drive efficient quantum simulation of quantum field theories.

For the results presented in this work, we implemented our own QITE algorithm for state evolution and QMETTS algorithm for computing thermal average using {\tt Python} with the {\tt Qiskit} library (version 0.44)~\cite{Qiskit}. Specifically, the unitary evolution is carried out with the {\tt PauliEvolutionGate} library, and the expectation are evaluated with the newly-released {\tt Sampler} class in {\tt Qiskit} to achieve the best possible performance on quantum simulators. In the numerical results, we used exact statevector simulation. Simulation with shot statistics and noise models to mimic realistic near-term devices can also be incorporated with our setup, although a much longer simulation time is expected.

\section{Free fermion fields at thermal equilibrium on qubits}\label{sec:free_fermion}
In this section, we focus on free fermion fields. In this case, the phase-space distribution $f_p$ for any given $N$ (and $a$) can be readily solved in momentum space. Moreover, the energy density and the pressure can be both expressed in terms of $f_p$. The main purpose of the section is to study the convergence of discrete results towards the continuum limit and to conduct quantum simulations, contrasting them with analytical expressions. 

\subsection{Thermal states of free fermion fields on lattice}

As $\hat{H}_0$ commutes with the momentum $\hat{p}$, the momentum eigenstate $|p\rangle_p$ is also the energy eigenstate. Using the anticommutation relations in eq.~(\ref{eq:anticommutingRelationForap}), one has
\begin{align}
    \hat{H}_0 |p\rangle_p = (E_p + E_{\Omega_F})|p\rangle_p,
\end{align}
where  $E_p$ given in eq.~(\ref{eq:Epas}) with its dependence on $m$ kept implicit, and $E_{\Omega_F}$ denotes the vacuum energy for the free fields, satisfying
\begin{align}
    \hat{H}_0|0\rangle_p = E_{\Omega_F}|0\rangle_p \equiv - \frac{1}{2}\sum_p E_p|0\rangle_p.
\end{align}
And, for free fields, one has
\begin{align}
    \psi_n(t) &= e^{i \hat{H}_0 t} \psi_n e^{- i \hat{H}_0 t} = \sum\limits_{n}\frac{1}{n!}\underbrace{[i\hat{H}_0,[i\hat{H}_0,[\cdots,[i\hat{H}_0}_{n},\psi_n]\cdots]]]\notag\\
    &= \sum\limits_p
    \frac{1}{\sqrt{2 N E_{p}}}[\hat{a}_p u_p e^{-i E_p t 
 + ipan}+\hat{a}_p^\dagger u^*_p e^{i E_p t -ipan}].
  \end{align}

By plugging the above equation into eq.~(\ref{eq:spectrualFunc}), one obtains the spectral function for the free fermions of the form
\begin{align}
    \rho_F(p^0, p) = 2\pi\text{sign}(p^0)(\slashed{p}+m)\delta((p^0)^2-E_p^2)
\end{align}
for $p\in\Omega_p$. As there is only one single-particle delta function in $\rho_F(p^0, p)$, the phase-space distribution function is accordingly defined as
\begin{align}
    f_p \equiv  \langle\hat{a}^\dagger_p \hat{a}_p\rangle_\beta.
\end{align}
In momentum space, one obtains straightforwardly
\begin{align}
    Z_\beta = e^{-\beta E_{\Omega_F}}\prod_{p} (1+e^{-\beta E_p}),\qquad \text{Tr}[ \hat{a}^\dagger_p \hat{a}_p e^{-\beta \hat{H}}] = e^{-\beta (E_p+E_{\Omega_F})}\prod_{p'\neq p} (1+e^{-\beta E_{p'}}),
\end{align}
yielding the Fermi-Dirac distribution
\begin{align}\label{eq:fFD}
    f_p=\frac{1}{e^{\beta E_p}+1}.
\end{align}

Equivalently, we define the phase-space distribution as
\begin{align}\label{eq:fFD_sim}
    f_p &= \frac{1}{2 E_p} \sum_n \big[\gamma^0 u_p\big]_{\alpha'}\big[ \bar{u}_p\gamma^0\big]_\alpha e^{ipan} \langle \bar{\psi}_{n}^{\alpha'}\psi_{0}^{\alpha}\rangle_\beta,
\end{align}
which is to be used in quantum simulations in the next subsection.
By using the relation between $\hat{a}_p$, $\hat{a}_p^\dagger$ and $\psi_n$ in eqs.~(\ref{eq:Psin}) and (\ref{eq:aAndPsi}), one can show that the above definition reduces to
\begin{align}
    f_p &= \sqrt{\frac{N}{2 E_p}}  \langle \hat{a}^\dagger_{p}[\bar{u}_p\gamma^0\psi_{0}]\rangle_\beta = \langle \hat{a}^\dagger_{p}\hat{a}_{p}\rangle_\beta,
\end{align}
where one needs to use the orthogonal relation $u^\dagger_p v_{-p}=0$ as well as the following relations 
\begin{align}\label{eq:aaExpect}
\langle \hat{a}^\dagger_{p}\hat{a}_{p'}\rangle_\beta=\delta_{pp'}\langle \hat{a}^\dagger_{p}\hat{a}_{p}\rangle_\beta,
~~~~\langle \hat{a}^\dagger_{p}\hat{a}^\dagger_{p'}\rangle_\beta= \delta_{p,-p'}\langle \hat{a}^\dagger_{p}\hat{a}^\dagger_{-p}\rangle_\beta,
~~~~\langle \hat{a}_{p}\hat{a}_{p'}\rangle_\beta= \delta_{p,-p'}\langle \hat{a}_{p}\hat{a}_{-p}\rangle_\beta.
\end{align}
These relations can be justified as follows. As $U_m$ commute with $\hat{H}$, one has
\begin{align}
0=\langle U_m^\dagger \hat{a}^\dagger_{p}U_m\hat{a}_{p'}\rangle_\beta -\langle \hat{a}^\dagger_{p}U_m\hat{a}_{p'} U_m^\dagger\rangle_\beta = (e^{-ipam}-e^{-ip'am}) \langle \hat{a}^\dagger_{p}\hat{a}_{p'}\rangle_\beta.
\end{align}
Therefore, the expectation value vanishes if $p\neq p'$, as required by momentum conservation. Similarly, one can conclude that the expectation values of $\hat{a}^\dagger_{p}\hat{a}^\dagger_{p'}$ and $\hat{a}_{p}\hat{a}_{p'}$ vanish if $p'\neq -p$.

Let us then derive the expression of the energy density in terms of $\hat{a}^\dagger_p$ and $\hat{a}_p$. Eliminating the time derivative of $\psi_n$ by using the Heisenberg equation of motion yields
\begin{align}
    \hat{\epsilon}_n - \hat{P}_n =& \frac{m}{2a}\bar\psi_{n} \psi_{n}-\frac{r}{4a^2}\bar\psi_{n}(\psi_{n+1}-2\psi_{n}+\psi_{n-1})=\frac{m}{a}\bigg(a_n^\dagger a_n -\frac{1}{2}\bigg)\notag\\
    & -\frac{r}{4a^2}[a_n^\dagger (a_{n+1} - 2 a_n + a_{n-1})-a_n (a^\dagger_{n+1} - 2 a^\dagger_n + a^\dagger_{n-1})]
\end{align}
with the pressure given by
\begin{align}
    \hat{P}_n =& - \frac{1}{4a^2}\bar{\psi}_n i\gamma^1 (\psi_{n+1}-\psi_{n-1})=-\frac{i}{4a^2}[a_n(a_{n+1}-a_{n-1}) + a^\dagger_n(a^\dagger_{n+1}-a^\dagger_{n-1})].    
\end{align}
In terms of the creation and annihilation operators $\hat{a}^\dagger_p$ and $\hat{a}_p$, after some algebra, one obtains
\begin{align}\label{eq:eF}
    \epsilon =\frac{1}{L}\sum_p {E_p}(\langle \hat{a}^\dagger_p \hat{a}_p\rangle - 1/2)=\frac{1}{L}\sum_p {E_p}(f_p - 1/2).
\end{align}
Here and below, the spatial location of $\epsilon$ is omitted. If one is only interested in the medium effects, the term corresponding to the factor of $1/2$ in the parentheses on the right-hand side of the equation above can be dropped as it is the same as the expectation value in the vacuum state. Here, the expectation values of $\hat{a}_p \hat{a}_{-p}$ and $\hat{a}^\dagger_p\hat{a}^\dagger_{-p}$ do not contribute as a result of the relations in eq.~(\ref{eq:aaExpect}) and energy conservation. The above expression of $\epsilon$ is simply the discretized version of that in the contiuum limit:
\begin{align}\label{eq:observable_cont}
    \bar{\epsilon} = \int\frac{dp}{2\pi}{\sqrt{p^2+m^2}} \bar{f}_p,
\end{align}
where $\bar{f}_p$ stands for the Fermi-Dirac distribution with the energy $E_p$ in $\bar{f}_p$ given by $\sqrt{p^2+m^2}$. Accordingly, the phase-space distribution in eq.~(\ref{eq:fFD}) can be viewed as the thermal fixed point in the real-time evolution of spatially homogeneous systems in the weak-coupling limit, similar to those for quarks in QCD kinetic theory~\cite{Kurkela:2018oqw, Kurkela:2018xxd, Du:2020zqg, Du:2020dvp, Cabodevila:2023htm}.

\subsection{Simulation results for free fermion fields}
\label{sec:simsFree}

While it is more efficient to simulate the fermion fields in momentum space due to the complete diagonalization of the Hamiltonian, quantum simulations become more efficient in coordinate space when interactions are considered, as interaction terms are typically nonlocal in momentum space~\cite{Jordan:2011ci}. Throughout this paper, we carry out simulations in coordinate space, aiming to compare them with the analytical results derived above in momentum space.

\begin{figure}
    \centering
    \subfigure[\label{fig:pressure_N4_r1} $T \gg m$ case
    ]{
    \includegraphics[width=0.45\textwidth]{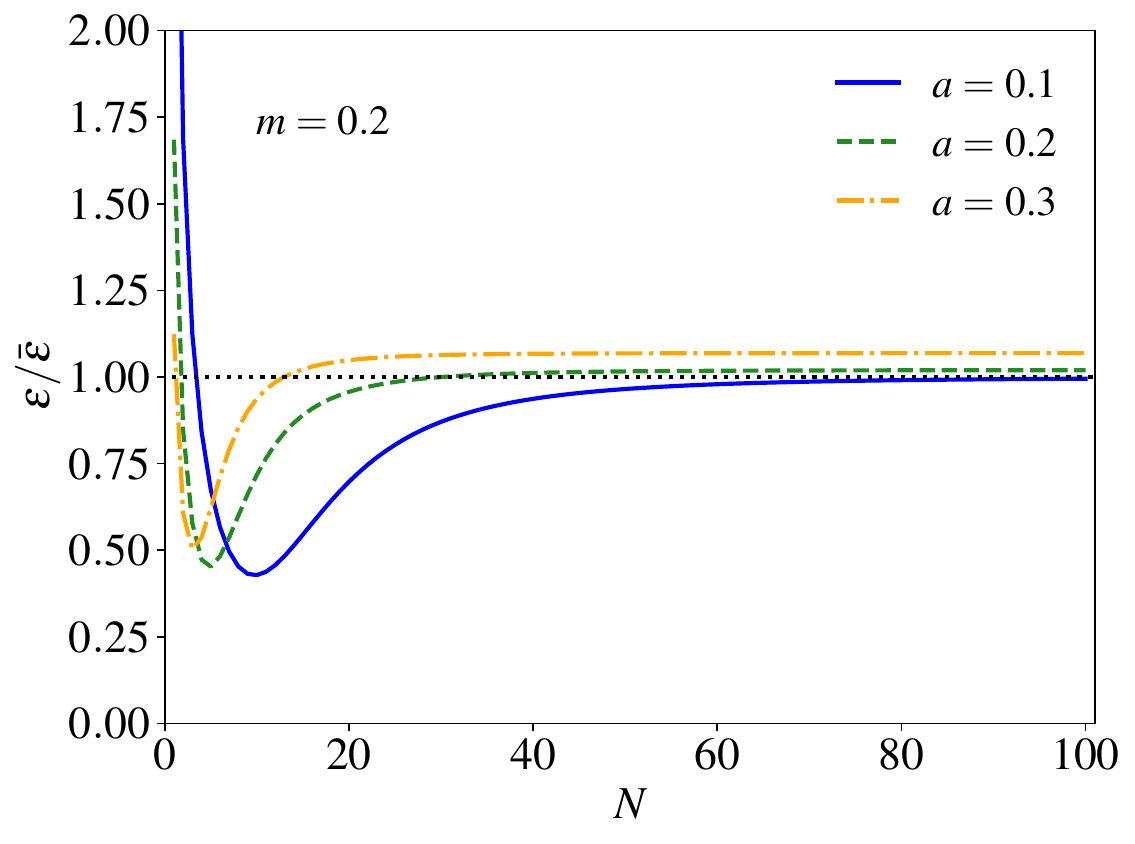}
    }
    \subfigure[\label{fig:energy_N4_r1} $m \gg T$ case
    ]{
    \includegraphics[width=0.45\textwidth]{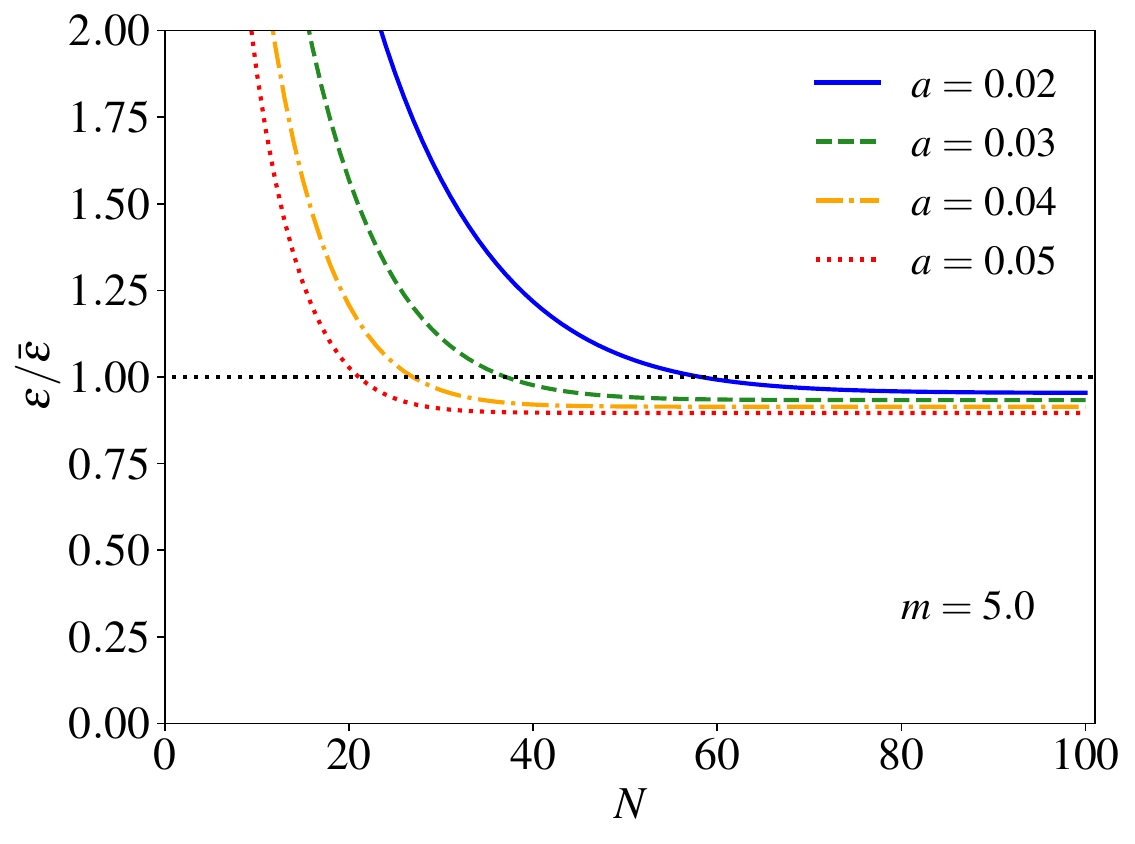}
    }
    \caption{Estimation of the number of qubits for the quantum simulation compared to the continuum [eq.~\eqref{eq:observable_cont}]. Both the small mass limit, $m=0.2$, and the large mass limit, $m=5.0$, of the lattice results are presented with a fixed temperature $T=1$, which sets the mass units.}
    \label{fig:lattice_to_continuum}
\end{figure}

Following the QITE algorithm introduced in sec.~\ref{sec:qite}, we numerically evaluate physical observables including the thermal distribution and energy density at various temperatures $T$ using the quantum simulator provided by {\tt Qiskit}. In thermal equilibrium, the energy of the particles is typically of order $T$. Accordingly, for $T\gg m$, one has the typical momentum $\bar{p}\sim T$; for $m \gg T$, one has $\bar{p}\sim\sqrt{2mT}$. Here, in the simulation, we discretize the position space as in eq.~(\ref{eq:lattice_discretization_x}), so for a typical simulation, we need the maximum momentum in the system $p_{max}=\pi/a > \bar{p}$ and the minimum momentum $\Delta p = 2\pi/(Na) < \bar{p}$, given the qubit number $N$ and the lattice spacing $a$. Ideally, one would want to use a very small spacing and a large number of qubits so that sufficient momentum modes are sampled in the range of interests. In fig.~\ref{fig:lattice_to_continuum}, we estimate the energy density $\epsilon$ in eq.~(\ref{eq:eF}) using the exact Fermi-Dirac distribution $f_p$ in eq.~(\ref{eq:fFD}) as a function of qubit number $N$ and compare them to the continuum limit for different lattice spacings. We study both the small mass limit, $m=0.2$, and the large mass limit, $m=5.0$, with a fixed temperature $T=1$, which establishes the mass units. We can see that potentially hundreds of qubits are necessary to recover the calculations at the continuum, along with appropriate lattice spacings. We can see that roughly 20 qubits are needed for the simulation to agree reasonably with the continuum (up to about 5-10\%) with lattice spacing $a=0.2$ for $m=0.2$ and $a=0.04$ for $m=5.0$, respectively. These choices of parameters are by no means optimal but serve as a starting point for our investigation with quantum simulation.

\begin{figure}
    \centering
    \subfigure[\label{fig:fp_N4_smallmass_r1} $T \gg m$ case
    ]{
    \includegraphics[width=0.45\textwidth]{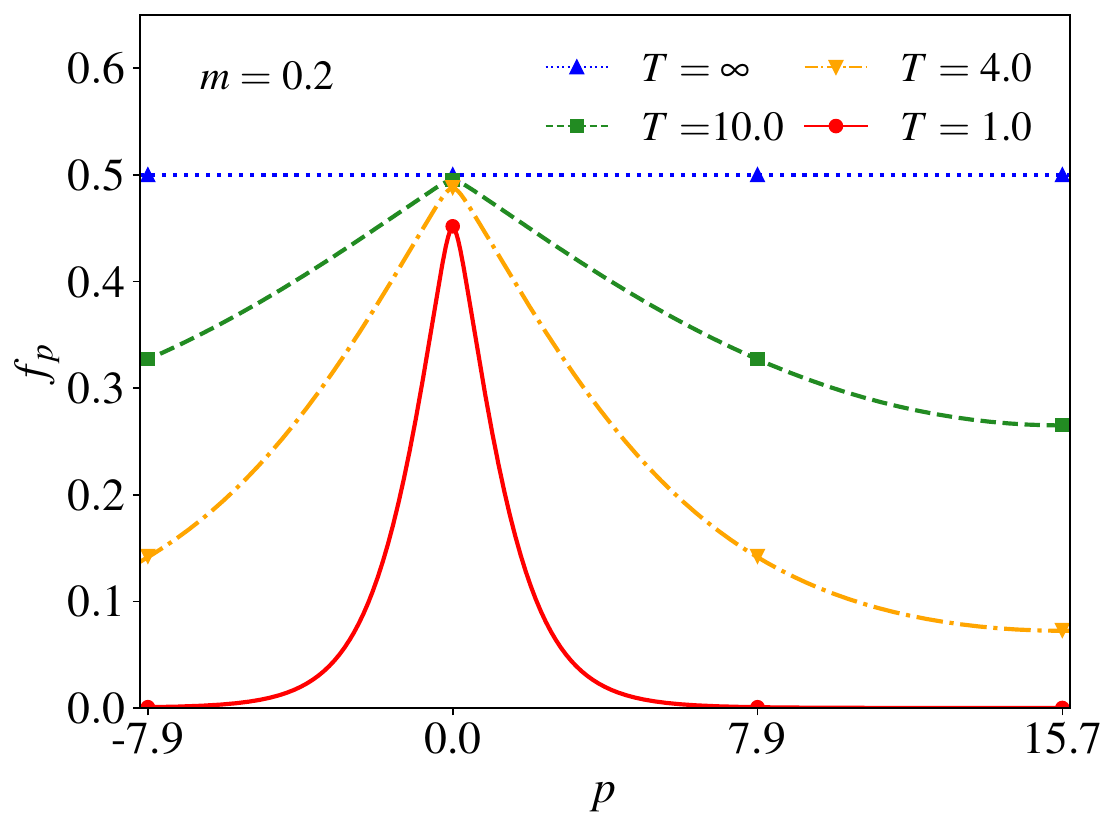}
    }
    \subfigure[\label{fig:fp_N4_largemass_r1} $m \gg T$ case
    ]{
    \includegraphics[width=0.45\textwidth]{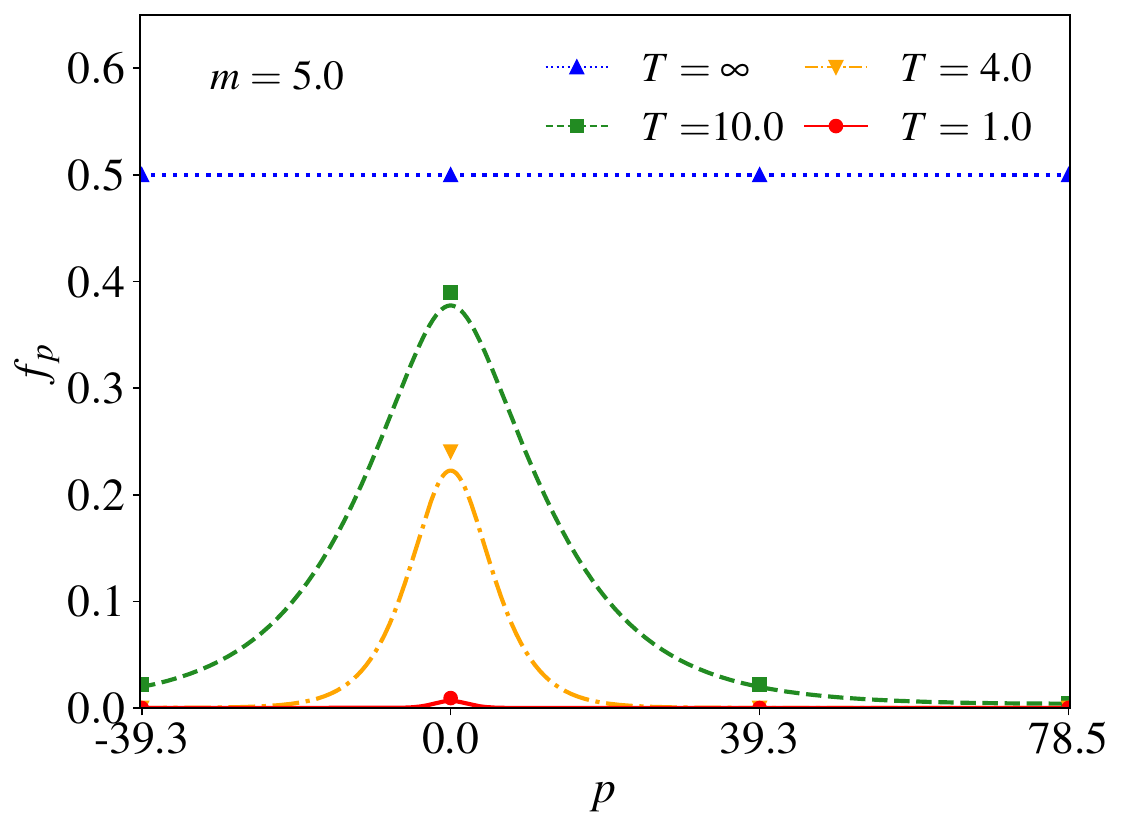}
    }
    \caption{Fermionic thermal distribution obtained from quantum simulation on 4 qubits in thermal limits (a) $T\gg m=0.2$ and (b) $m=5.0\gg T$. Simulation results are in solid markers at discretized momenta using the smallest trotterization step $\Delta \beta=0.001$. Analytical lines of Fermi-Dirac distributions are provided [eq.~(\ref{eq:fFD})] for comparison.}
    \label{fig:fp}
\end{figure}

In fig.~\ref{fig:fp}, we show the quantum simulation results of the thermal distribution $f_p$ [eq.~(\ref{eq:fFD_sim})] of the free fields in the small and large mass limits. In both cases, the free fermion Hamiltonian is evolved with the QITE algorithm from an initial temperature $T =\infty$ ($\beta=0$) to the final temperature $T=1$. The Wilson parameter $r=1$ is used throughout this work. Estimations of $f_p(T)$ along the trajectory are provided at selected $T=\infty, 10.0, 4.0$, and $1.0$. Despite the limited number of momentum modes, we can see that at both mass limits, the simulated results of $f_p$ are in good agreement with the analytical curves, i.e., Fermi-Dirac distributions in eq.~\eqref{eq:fFD}. Specifically, we used increasingly smaller trotterization steps of $\Delta \beta= 0.02, 0.01, 0.002, 0.001$ such that the QITE simulation converges at the targeted temperature $T=1$. Results of the smallest $\Delta \beta = 0.001$ are presented. Notably, simulation to the larger mass is more sensitive to the trotterization step, as evidenced by the minor discrepancies between the analytical and simulation lines in fig.~\ref{fig:fp_N4_largemass_r1}.

\begin{figure}
    \centering
    \subfigure[\label{fig:energy_N4_r1} $T \gg m$ case
    ]{
    \includegraphics[width=0.44\textwidth]{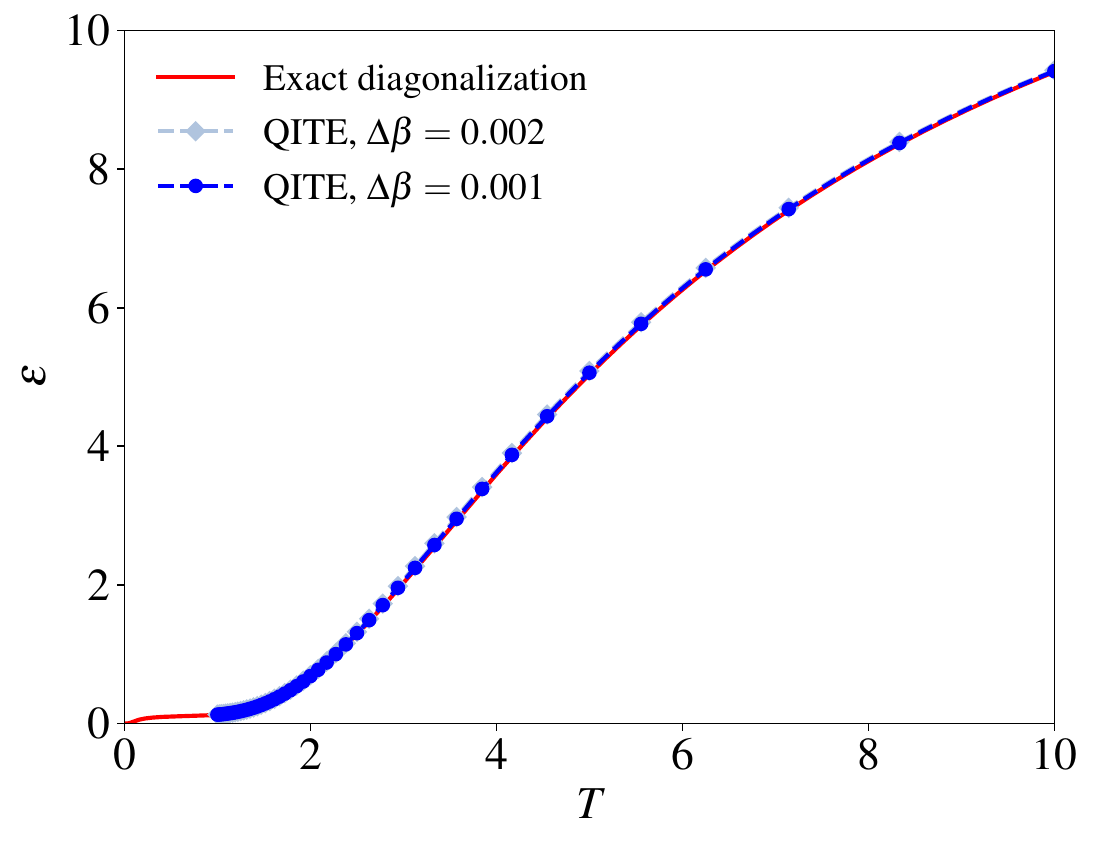}
    }     \subfigure[\label{fig:energy_N4_r1} $m \gg T$ case
    ]{
    \includegraphics[width=0.44\textwidth]{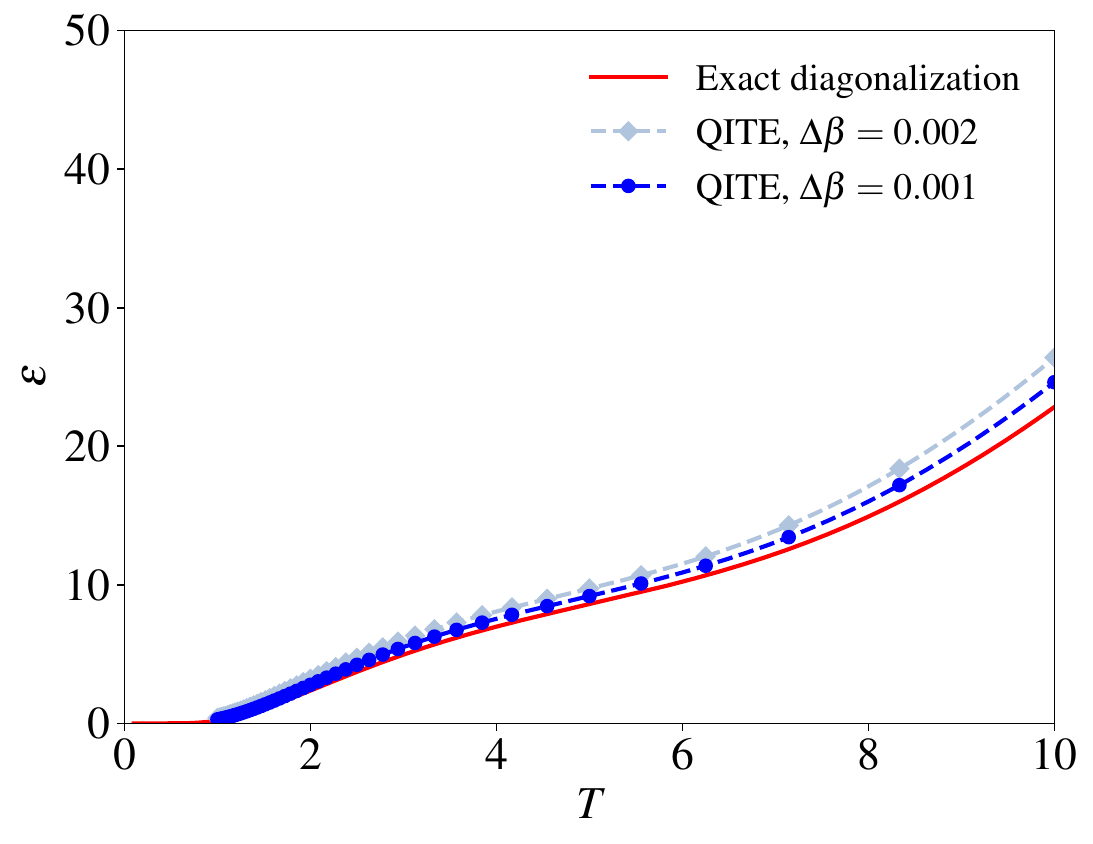}
    }
    \caption{Fermionic energy density $\epsilon$ obtained from quantum simulation on 4 qubits in thermal limits (a) $T\gg m=0.2$ and (b) $m=5.0\gg T$. Two trotterizations of $\Delta \beta$ = 0.001 (0.002) are presented by the dark (light) colored lines in the QITE evolution to $T=1$. In comparison, calculations using exact diagonalization are presented by the solid red lines.}
    \label{fig:pressure_and_energy}
\end{figure}

Likewise, we calculate the energy density $\epsilon$ of the same free fermion fields with the QITE algorithm and present the results in fig.~\ref{fig:pressure_and_energy}. For numerical accuracy, we also provide the exact estimation of $\epsilon$ by evaluating their traces via diagonalization of the Hamiltonian. We can see that with decreasing trotterization step $\Delta \beta$ the quantum simulation results using the QITE algorithm gradually approach the exact diagonalization for both $m=0.2$ and $m=5.0$. Note that smaller trotterization steps are needed for the large mass limit, which is also suggested in the simulation results of $f_p$. 

\section{Interacting fermion fields at  thermal equilibrium on qubits}\label{sec:interaction_fermion}
In this section, we investigate fermionic systems coupled through four-fermion interactions. Unlike Dirac fermions,  the four-fermion interactions involving identical Majorana fields effectively vanish due to the Pauli exclusion principle. One, hence, needs to introduce another Majorana field.\footnote{If this additional Majorana field has the same mass and the same space-time dependence as $\psi$, one effectively recovers the Gross-Neveu model~\cite{Gross:1974jv} (with the number of Dirac fermion fields equal to 1), as studied previously in \cite{Jordan:2014tma, Czajka:2021yll}.}

To minimize resource requirements for quantum simulations and make the model solvable analytically, the additional Majorana field, denoted as $\psi_B$, is assumed to be homogeneous in space and does not vary with $x$ below. Accordingly, the Lagrangian takes the form
\begin{align}\label{eq:Lint}
    L = \int dx\bigg[\frac{1}{2}\bar{\psi}(i\slashed{\partial}-m_0)\psi  - \frac{g}{4}(\bar\psi\psi) (\bar{\psi}_B\psi_B)\bigg] + \frac{1}{2}\bar{\psi}_B(i\gamma^0\partial_t-M)\psi_B,
\end{align}
where the mass $m_0$, $M$ and $g$ are all bare quantities, and the field $\psi_B$ satisfies the following anticommutation relations
\begin{align}
\{\psi_B, \psi_B\}=1,\qquad \{\psi, \psi_B\}=0.    
\end{align}
In this case, the interaction Hamiltonian takes the form
\begin{align}
     \hat{H}_I = \frac{M}{2}\bar{\psi}_B\psi_B + \frac{g}{4}\int dx\bar\psi\psi \bar{\psi}_B\psi_B.   
\end{align}
As it only has the zero-momentum mode, $\psi_B$ can be expressed in the form
\begin{align}\label{eq:psiM}
    \psi_B & \equiv 
    \frac{1}{\sqrt{2 M}}[ u_0(M) {b} + u_0^*(M) {b}^\dagger]=\frac{1}{\sqrt{2}}\begin{pmatrix}
    1&1\\
    i&-i
\end{pmatrix}
\left(
    \begin{array}{c}
    b      \\
     b^\dagger
    \end{array}
    \right),
\end{align}
where ${b}^\dagger$ and ${b}$ are respectively the creation and annihilation operators of $\psi_B$, and the dependence of $u_p$ on mass $M$, as defined in eq.~(\ref{eq:spinors}), is explicitly indicated as $u_p(M)$.

\subsection{The continuum theory and renormalization}
In 1+1 dimensions, the QFT in eq.~(\ref{eq:Lint}) is renormalizable. This can be easily concluded from the mass dimensions of its parameters. By utilizing the fact that the mass dimension of $L$, denoted by $[L]$, equals 1, one has
\begin{align}
    [\psi]=\frac{1}{2},\qquad[\psi_B]=0,\qquad [g] = [m_0] = [M] = 1.
\end{align}
That is, the theory is in fact super-renormalizable.

Let us now analyze the behavior of Green functions at short distances. For this, one needs the free propagator of $\psi_B$, which, according to eq.~(\ref{eq:psiM}), takes the form
\begin{align}
    \langle0|\text{T}\psi_B(t) \bar\psi_B(0)|0\rangle = \int\frac{d\omega}{2\pi}\frac{i(\omega\gamma^0 + M)}{\omega^2 - M^2 + i\epsilon}e^{-i\omega t}.
\end{align}
That is, it is similar to the free propagator of $\psi$, except that its internal momentum only has the zero component. Then, it is straightforward to obtain the superficial degree of divergence, denoted as $D$, for a generic diagram $G$ containing $V$ vertices, $N_{\psi}$ external lines of $\psi$, and $N_{B}$ external lines of $\psi_B$. Such a graph contains $(2V-N_{\psi})/2$ free propagators of $\psi$ and $(2V-N_B)/2$ free propagators of $\psi_B$. Here, each free propagator of $\psi$ or $\psi_B$ adds 1 or 0 to $D$, respectively. Considering additionally that each vertex has a delta function in 2-momentum with one enforcing overall 2-momentum conservation, the superficial degree of divergence reads
\begin{align}
    D = -2(V-1) + (2V - N_{\psi})/2 = 2 -V - N_{\psi}/2.
\end{align}

Let us focus on $\psi$. For $\psi$, the only diagram potentially UV divergent is the self-energy at one loop, which has $D =0$.  However, one can see that the one-loop diagram is in fact UV finite as a consequence of $\text{Tr}(\gamma^0)=0$. Therefore,  all the diagrams containing at least two external lines of $\psi$ are UV finite. Accordingly, the renormalization constants involving $\psi$ in this theory are finite, as a consequence of eliminating high-momentum modes by making $\psi_B$ homogeneous in space. This is qualitatively different from the Gross-Neuve model.

\subsection{The discrete theory and quasiparticles}
In this subsection, we show that the Hamiltonian in the discrete theory can be diagonalized and, hence, the model is solvable analytically.

\subsubsection{Diagonalization of the discrete Hamiltonian}

Discretizing $\psi$ as in sec.~\ref{sec:discretization} and expressing $\psi_B$ in terms of $b^\dagger$ and $b$ in the continuum Hamiltonian yields
\begin{align}\label{eq:HInt}
    \hat{H} =& \hat{H}_0 + M \bigg(b^\dagger b-\frac{1}{2}\bigg) + g\sum_n \bigg(a_n^\dagger a_n-\frac{1}{2}\bigg)\bigg(b^\dagger b-\frac{1}{2}\bigg),
\end{align}
where $\hat{H}_0$ is given in eq.~(\ref{eq:H0}) with free mass $m$ replaced by the bare mass $m_0$. We work in coordinate space and choose the complete orthonormal basis as
\begin{align}
    &|n_0 n_1\cdots n_{N-1} n_N \rangle_x\equiv (a^\dagger_0)^{n_0}(a^\dagger_1)^{n_1}\cdots(a^\dagger_{N-1})^{n_{N-1}} (b^\dagger)^{n_N} |0\rangle_x,\notag\\
    &{}_x\langle n_0 n_1\cdots n_{N-1} n_N|\equiv {}_x\langle 0 |(b)^{n_N}(a_{N-1})^{n_{N-1}}(a_{N-2})^{n_{N-2}}\cdots(a_0)^{n_0},
\end{align}
where $n_i=0$ or $1$, and the vacuum state in coordinate space $|0\rangle_x$ is defined by $b|0\rangle_x=0$ and $a_n|0\rangle_x=0$ for all $n\in\Omega_x$. Again, we use the Jordan-Wigner transformation to map the $N+1$ fermions to $N+1$ qubits in quantum simulations.

In this discrete theory, there are five parameters: $a$, $N$, $m_0$, $M$ and $g$. Among them, $a$ and $N$ are input parameters. In addition, one needs three renormalization conditions in order to fix the values of the three bare parameters: $m_0$, $M$ and $g$ (at scale $a$) in relation to the renormalized ones. As we focus on the effects of interactions in comparison with the free theory, we only implement the mass renormalization of $\psi$ in order to introduce the physical mass $\bar{m}$, which is taken to be equal to the mass for the free field studied in the previous section. As detailed below, $M$ is assumed to be heavy. In this case, for the mass renormalization of $\psi$, one only needs to evaluate the lowest two eigenstates of $\hat{H}$, with their eigenvalues denoted respectively by $E_{\Omega}$ and $E_0$: the physical vacuum state $|\Omega\rangle$ and the mass eigenstate. The mass renormalization is implemented as follows~\cite{Jordan:2014tma}
\begin{align}
    \bar{m}= E_0 - E_{\Omega}
\end{align}
with $\bar{m}$ the physical (renormalized) mass of the field $\psi$.

In the above representation, $\hat{H}$ is a Hermitian $2^{N+1}\times 2^{N+1}$ matrix. All its off-diagonal matrix elements vanish between the eigenstates of $b^\dagger b$, which is evident in the expression of the Hamiltonian in eq.~(\ref{eq:HInt}). This is a consequence of momentum conservation among the two $\psi$ fields at each vertex. Consequently, energy is also conserved among them and the transition between the two energy levels of $\psi_B$ is forbidden. Accordingly, the Hamiltonian can be divided into the following blocks
\begin{align}
    \hat{H} = \left(
    \begin{array}{cc}
       \hat{H}_{\psi}^0  & 0 \\
        0 & \hat{H}_{\psi}^1    \end{array}
    \right).
\end{align}
Here, $\hat{H}_{\psi}^0$ is a $2^N\times 2^N$ block matrix corresponding to the state vector for $\hat{b}b=0$, which can be written as
\begin{align}
    \hat{H}_{\psi}^0 &= \hat{H}_0 - \frac{M}{2} -\frac{g}{2}\bigg(a_n^\dagger a_n-\frac{1}{2}\bigg)=  \sum\limits_p E_p(\bar{m}) \bigg(\hat{a}^\dagger_p \hat{a}_p - \frac{1}{2}\bigg)-\frac{M}{2},
\end{align}
where $\bar{m}\equiv m_0 - g/2$ and the creation and annihilation operators $\hat{a}_p^\dagger$ and $\hat{a}_p$ are related to $a^\dagger_n$ and $a_n$ according to eq.~(\ref{eq:ahat_p}) with $m$ replaced by $\bar{m}$. Similarly, for $\hat{b}b=1$, one has
\begin{align}
    \hat{H}_{\psi}^1 &= \hat{H}_0 + \frac{M}{2} + \frac{g}{2}\bigg(a_n^\dagger a_n-\frac{1}{2}\bigg) = \sum\limits_p E_p(\bar{m}+g) \bigg(\hat{a}'^\dagger_p \hat{a}'_p - \frac{1}{2}\bigg)+\frac{M}{2},
\end{align}
 where $\hat{a}_p'^\dagger$ and $\hat{a}'_p$ are related to $a^\dagger_n$ and $a_n$ through the same transformations as $\hat{a}^\dagger_p$ and $\hat{a}_p$ above except that $m$ is now replaced by $\bar{m}+g$ in the transformations. Note that $\hat{a}_p^\dagger\hat{a}_p$ and $\hat{a}_p'^\dagger{\hat{a}}'_p$ are so defined that they vanish when acting on eigenstates of $b^\dagger b$ with eigenvalues of 1 and 0, respectively.

 In the following discussions, $g$ and $M$ are both taken to be positive. In this case, the lowest two eigenvalues are given by
\begin{align}
    E_{\Omega} = -\frac{1}{2}\sum\limits_p E_p(\bar{m}) - \frac{M}{2}, \qquad E_0 = \bar{m} + E_{\Omega},
\end{align}
yielding the renormalized mass $\bar{m} \equiv m_0 - g/2$.

\subsubsection{Quasiparticles in thermal systems}
The doubling of the energy levels in this model gives arise to an additional type of quasiparticles in thermal systems. Let us illustrate this point with $N=1$. In this case, one has
\begin{align}
    \hat{H} =& \bar{m}\bigg(a_0^\dagger a_0-\frac{1}{2}\bigg) + m_B \bigg(b^\dagger b-\frac{1}{2}\bigg) + g a_0^\dagger a_0 b^\dagger b - \frac{g}{4}
\end{align}
with the renormalized mass $\bar{m} =  m_0 - {g}/{2}$ and the background field mass $m_B\equiv M - {g}/{2}$. It is easy to see that the full Hamiltonian shares the same eigenstates with the free Hamiltonian albeit the energy eigenvalues have been modified by the interactions:
\begin{align}
    E - E_{\Omega} = 0, \bar{m}, m_B, \bar{m} + m_B + g,
\end{align}
where the vacuum energy $E_{\Omega} = -\bar{m}/2 - M/2$, and the corresponding eigenstates are given by direct products of the eigenstates of $a_0^\dagger a_0$ and $b^\dagger b$:
$|00\rangle, |10\rangle, |01\rangle, |11\rangle$ with the physical vacuum state $|0\rangle$ defined by $a_0|0\rangle=0 =b|0\rangle$. 

In order to calculate the spectral density in thermal systems, we express the evolution operator in the form
\begin{align}
    e^{-i \hat{H} t + i E_{\Omega} t} = |00\rangle\langle 00| + e^{-i m_B t} |01\rangle\langle 01| + e^{-i \bar{m} t}|10\rangle\langle 10| + e^{-i (\bar{m} + m_B + g)t}|11\rangle\langle 11|.
\end{align}
The time evolution of $a_0$ is then given by
\begin{align}
a_0(t) = e^{i\hat{H}t} a_0 e^{-i\hat{H}t} = e^{-i \bar{m} t} |00\rangle \langle 10| + e^{-i (\bar{m} + g) t} |01\rangle \langle 11|.
\end{align}
Accordingly, one can show that
\begin{align}
    \{\psi(t), \bar{\psi}\} &= [\cos(\bar{m} t) (|00\rangle\langle 00| + |10\rangle\langle 10|) + \cos((\bar{m} + g)t) (|01\rangle\langle 01| + |11\rangle\langle 11|)]\gamma^0\notag\\
    &-i[ \sin(\bar{m} t) (|00\rangle\langle 00| + |10\rangle\langle 10|) +  \sin((\bar{m} + g)t) (|01\rangle\langle 01| + |11\rangle\langle 11|)].
\end{align}
After taking thermal average, the spectral function is given by
\begin{align}
\rho_A(p^0) =& \int dt e^{ip^0 t} \langle \{\psi(t), \bar{\psi}\} \rangle_\beta \notag\\
 = & 2\pi\, \text{sign}(p^0) (p^0\gamma^0 + \bar{m})\delta((p^0)^2 - \bar{m}^2) (1-P_{B})\notag\\
& + 2\pi\, \text{sign}(p^0) (p^0\gamma^0 + \bar{m} + g)\delta((p^0)^2 - (\bar{m}+g)^2)P_B,
\end{align}
where the probability $P_B$ is defined as
\begin{align}
    P_B = Z_\beta^{-1} e^{-\beta E_{\Omega}^1} (1 + e^{-\beta(\bar{m} + g)})
\end{align}
with $E_{\Omega}^1 = -\bar{m}/2 - g/2 + M/2$, and 
\begin{align}
    Z_\beta = (1 + e^{-\beta \bar{m}} + e^{-\beta m_B} + e^{-\beta (\bar{m} + m_B + g)})e^{-\beta E_{\Omega}}.
\end{align}
That is, the presence of $\psi_B$ generates an additional type of quasiparticles with a different mass of $\bar{m}+g$.

\subsubsection{Thermal states of interacting fermion fields on lattice}

Given the diagonalized Hamiltonian above, this model is solvable analytically. The partition function is given by
\begin{align}
    Z_\beta = Z_{\beta}^0 + Z_{\beta}^1,
\end{align}
where the contributions to $Z_\beta$ respectively from $H_{\psi}^0$ and $H_{\psi}^1$ take the form 
\begin{align}
    Z_\beta^0 \equiv e^{-\beta E_{\Omega}}\prod_{p} (1+e^{-\beta E_p(\bar{m})}),\qquad Z_\beta^1\equiv e^{-\beta E_{\Omega}^1}\prod_{p} (1+e^{-\beta E_p(\bar{m}+g)})    
\end{align}
with  the lowest energy level of $H_{\psi}^1$ denoted as
\begin{align}
    E_{\Omega}^1 = -\frac{1}{2}\sum\limits_p E_p(\bar{m}+g) + \frac{M}{2}.
\end{align}
As illustrated by the case with $N=1$, there are two types of quasiparticles in thermal systems. Accordingly, we define two phase-space distributions in terms of $\hat{a}^\dagger_p \hat{a}_p$ and $\hat{a}'^\dagger_p \hat{a}'_p$. After some algebra, one has
\begin{align}\label{eq:fp_int}
    f^0_p \equiv \langle \hat{a}^\dagger_p \hat{a}_p\rangle_\beta = \frac{Z_\beta^0}{Z_\beta}\frac{1}{1+e^{\beta E_p(\bar{m})}},\qquad
    f^1_p \equiv \langle \hat{a}'^\dagger_p \hat{a}'_p\rangle_\beta = \frac{Z_\beta^1}{Z_\beta}\frac{1}{1+e^{\beta E_p(\bar{m}+g)}}.
\end{align}
Notably, the number operators $\hat{a}^\dagger_p \hat{a}_p$ and $\hat{a}'^\dagger_p \hat{a}'_p$ act on their exclusive subspaces $\hat{H}_\psi^0$ and $\hat{H}_\psi^1$, respectively, and they are rewritten in terms of the fermion fields $a^\dagger_n$ and $a_n$ in practical simulation.

In terms of the two phase-space distributions of quasiparticles, one can evaluate thermal properties of the system. As a consistency check, let us evaluate the energy density in the system. According to eq.~(\ref{eq:spatialHomo}), one has $\epsilon = \langle \hat{H} \rangle_\beta/L$. Then, it reduces to taking trace in subspaces $\hat{H}_\psi^0$ and $\hat{H}_\psi^1$:
\begin{align}\label{eq:ene_density_int}
   \epsilon =\frac{1}{Z_\beta L}[\text{Tr}(\hat{H}^0_{\psi} e^{-\beta\hat{H}^0_{\psi}})+\text{Tr}(\hat{H}^1_{\psi} e^{-\beta\hat{H}^1_{\psi}})] = \frac{Z_\beta^0}{Z_\beta} \epsilon^0 +  \frac{Z_\beta^1}{Z_\beta} \epsilon^1% = \frac{1}{L}\sum_p[ E_p(m)f_p^0 + E_p(m+g)f_p^1] + \epsilon|_{T\to0},
\end{align}
with
\begin{align}
\epsilon^0\equiv \frac{1}{L}\bigg[\sum_p \frac{E_p(\bar{m})}{1+e^{\beta E_p(\bar{m})}} + E_{\Omega}\bigg],\qquad 
\epsilon^1\equiv \frac{1}{L}\bigg[\sum_p \frac{E_p(\bar{m}+g)}{1+e^{\beta E_p(\bar{m}+g)}} + E_{\Omega}^1\bigg].
\end{align}
Note that in the limit $M\to\infty$, ${Z_\beta^1}/{Z_\beta}$ vanishes. Consequently, thermal properties of the system become identical to those in the free field theory, and the presence of $\psi_B$ only renormalizes the mass of $\psi$.

\subsection{Simulation results for interacting fermion fields}
\begin{figure}
    \centering
    \subfigure[\label{fig:fpInt_small_m} $T \gg \bar{m}$ case with $\bar{m} = 0.2, M=10.0$
    ]{\includegraphics[width=0.48\textwidth]{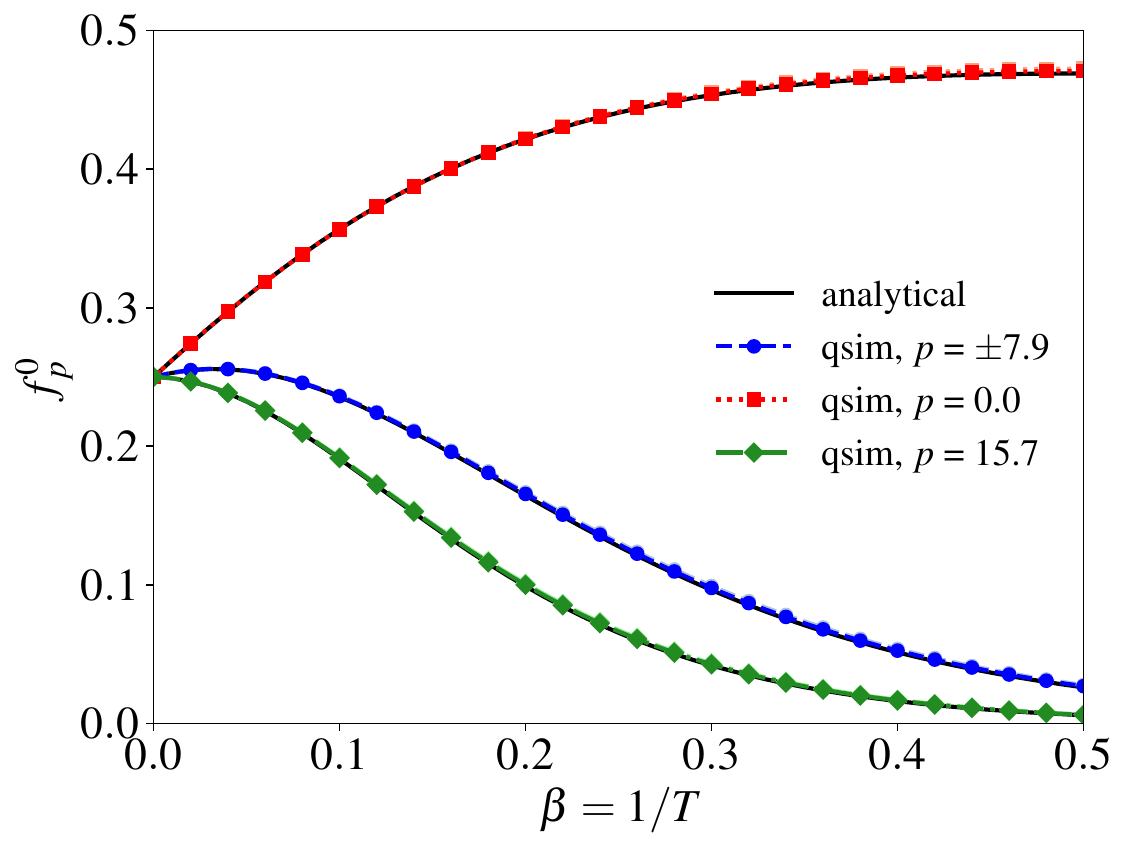}
    \includegraphics[width=0.48\textwidth]{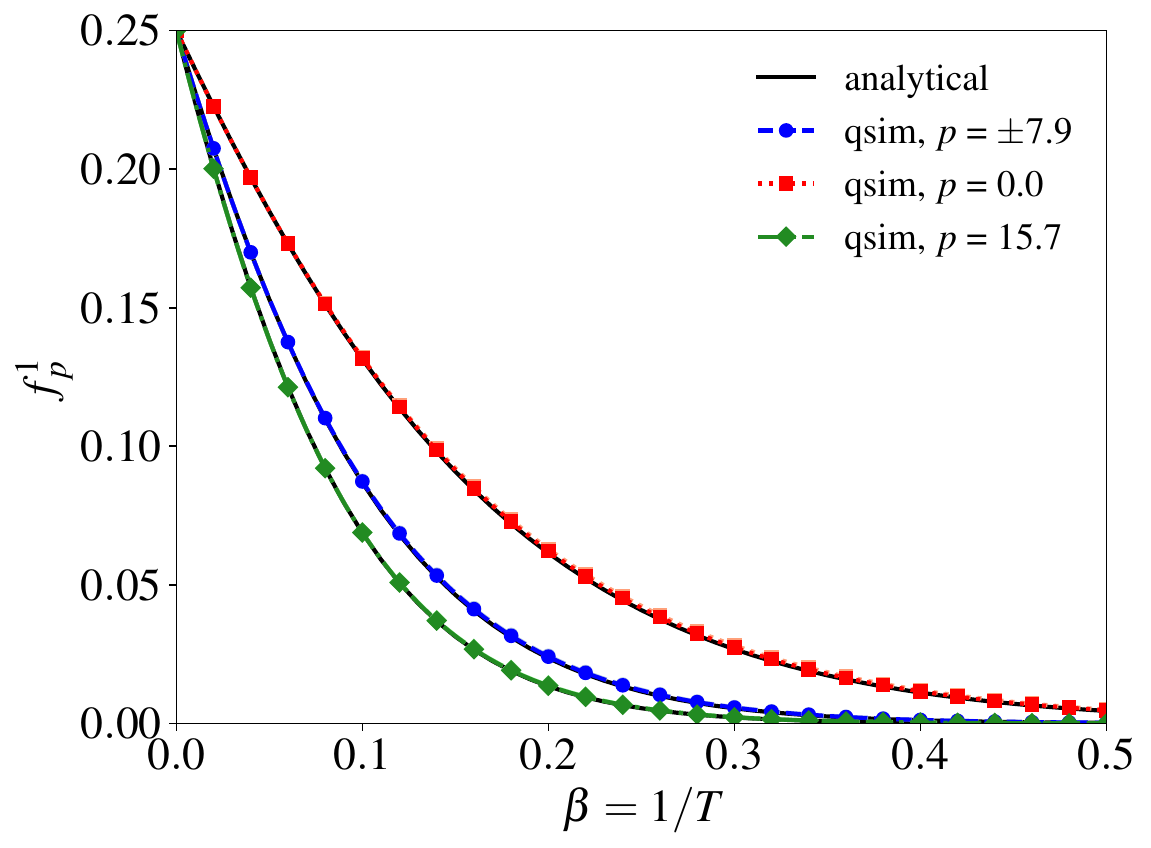}}
    
    \subfigure[\label{fig:fpInt_large_m} $\bar{m} \gg T$ case with $\bar{m} = 5.0, M=10.0$
    ]{\includegraphics[width=0.48\textwidth]{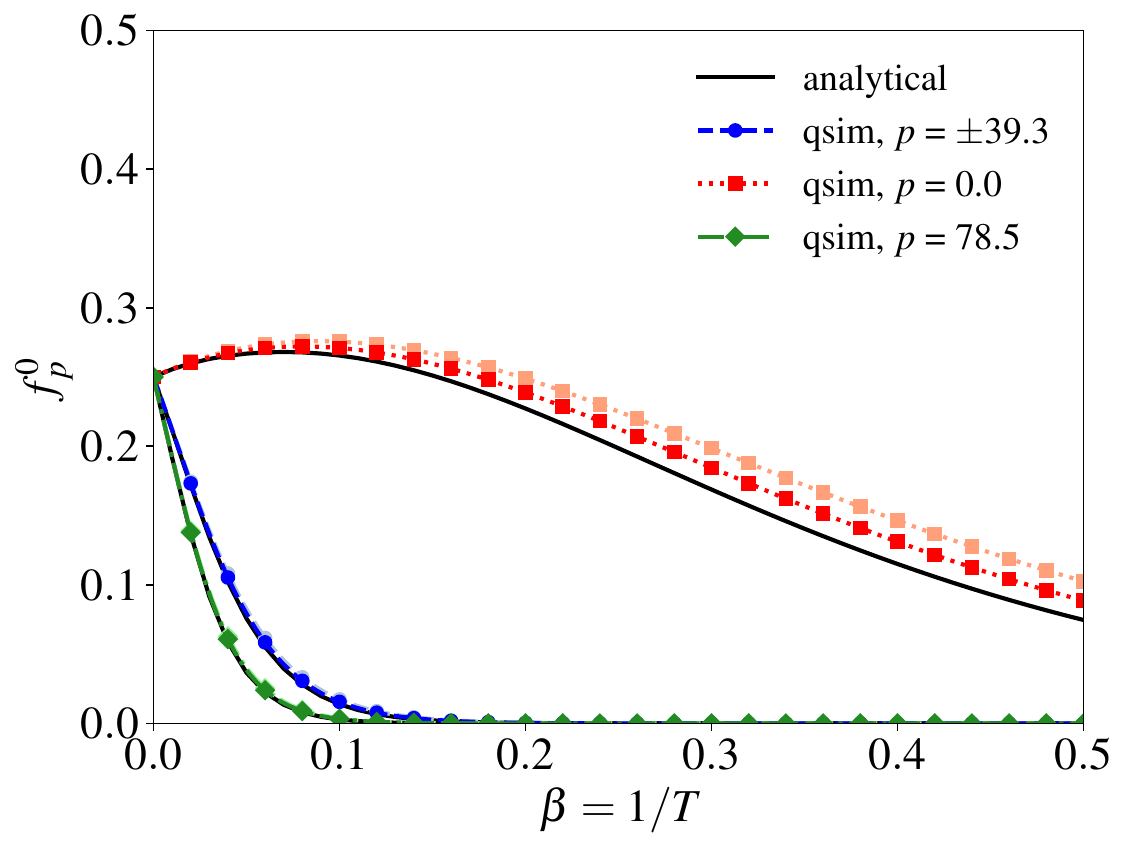}
    \includegraphics[width=0.48\textwidth]{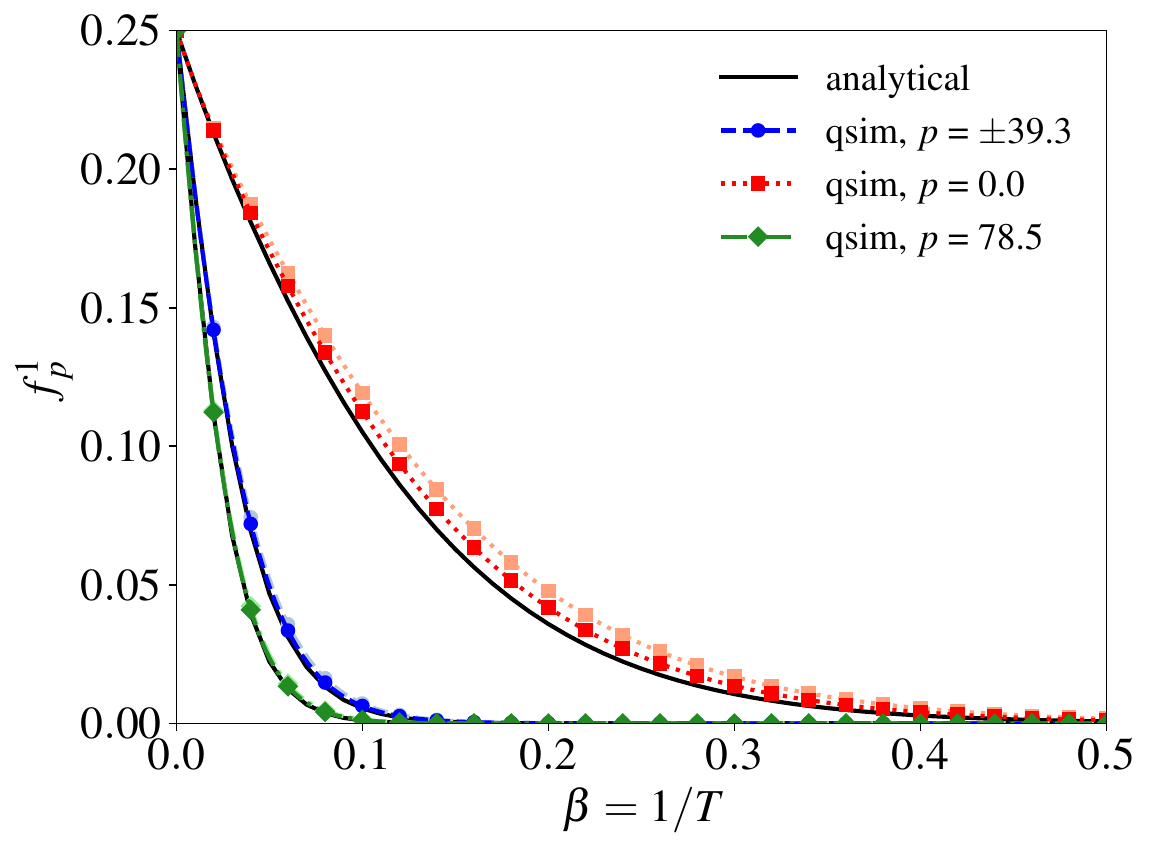}}
    \caption{Quantum simulation of thermal distribution functions $f_p^0$ and $f_p^1$ for each type of quasiparticles of the interacting fields with $N=4$ qubits for $\psi$ and one qubit for $\psi_B$. Two trotterizations of $\Delta \beta$ = 0.001 (0.002) are presented by the dark (light) colored lines. Analytical results (solid black lines) are calculated using eq.~\eqref{eq:fp_int}. Note that the thermal distributions for $p=\pm7.85$ or $p=\pm39.27$ are the same  and different vertical axes are used for $f_p^0$ and $f_p^1$.}
    \label{fig:fp_Hint}
\end{figure}

\begin{figure}
    \centering
    \subfigure[\label{fig:fpInt_large_m} $T \gg \bar{m}$ case]{\includegraphics[width=0.45\textwidth]{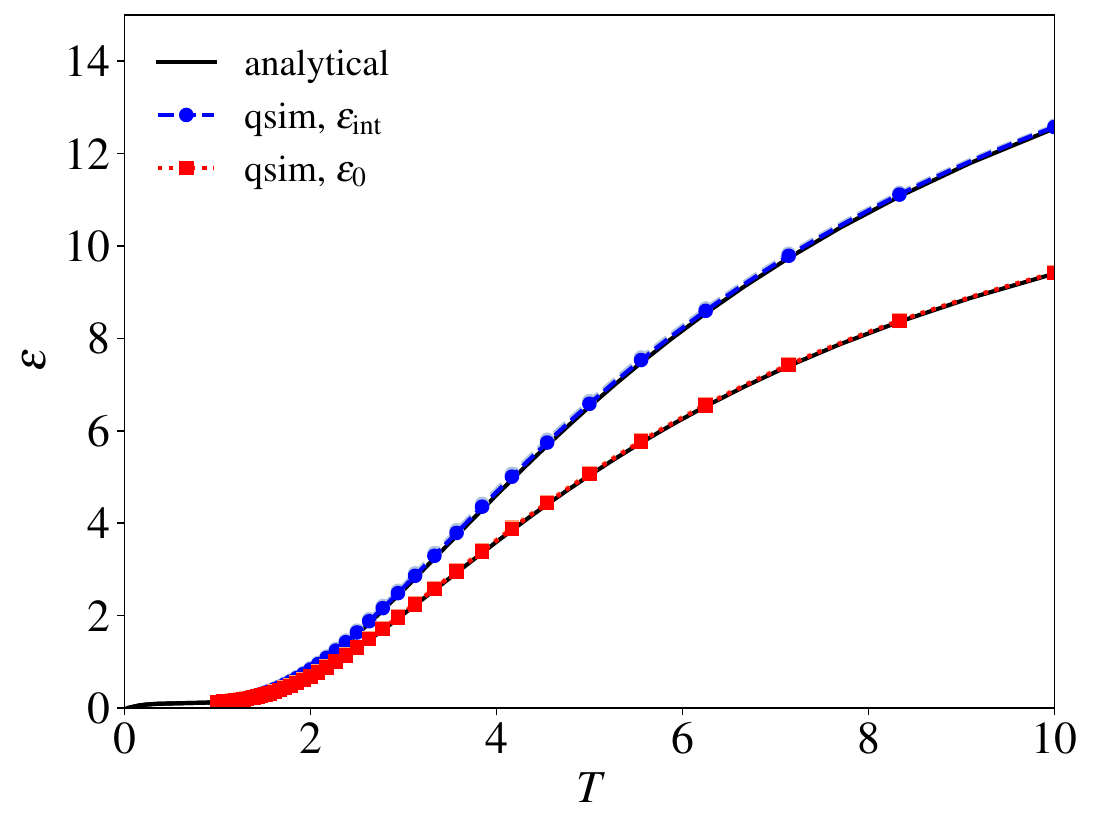}}
    \subfigure[\label{fig:fpInt_large_m} $\bar{m} \gg T$ case]{\includegraphics[width=0.45\textwidth]{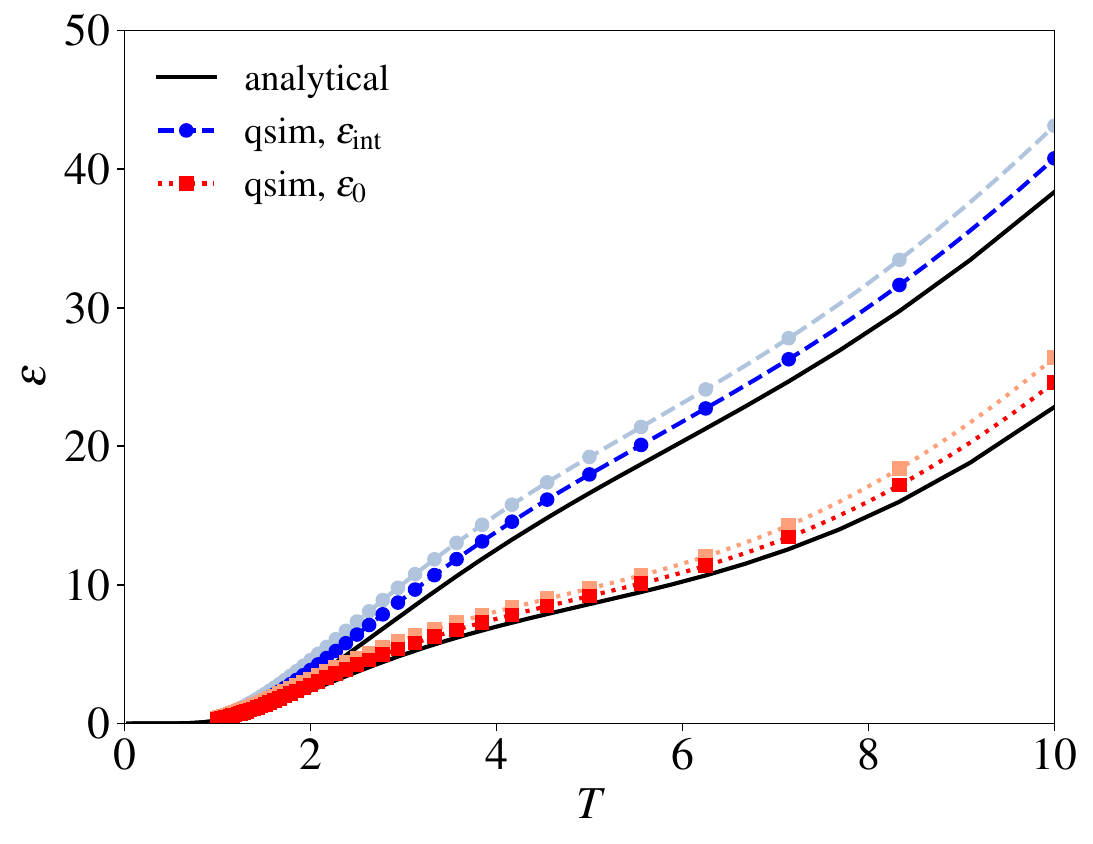}}
    \caption{Quantum simulation of the energy density for the interacting fields ($\epsilon_\mathrm{int}$) compared with the free fields ($\epsilon_0$) using $N=4$ qubits for $\psi$ and one qubit for $\psi_B$. Two trotterizations of $\Delta \beta$ = 0.001 (0.002) are presented by the dark (light) colored lines. Analytical results (solid black lines) are calculated using eq.~\eqref{eq:ene_density_int}.}
    \label{fig:e_Hint}
\end{figure}

In this section, we study how the results in sec.~\ref{sec:simsFree} are modified by the presence of the background field $\psi_B$. Specifically, we study the interacting fields with $N=4$ qubits for $\psi$ and one qubit for $\psi_B$ using the QITE algorithm. In the small mass and large mass limits for $\psi$, we choose the renormalized mass $\bar{m}=0.2$ and $\bar{m}=5.0$, respectively, such that they are the same as the values of $m$ used in the free field theory. For the background field $\psi_B$, we take $M=10.0$ (using the same units as those in sec.~\ref{sec:simsFree}) such that it is always heavier than $\bar{m}=m$ as our focus lies on the modification to the original free field $\psi$. Lastly, we take the coupling strength $g=1$. These choices of parameters serve as a starting point to investigate the interacting field theory of Majorana fermion fields.

We now show the simulation results of the thermal distributions and the energy density for the interacting fermion fields. In fig.~\ref{fig:fp_Hint}, we present the thermal distributions $f_p^0$ and $f_p^1$ for the quasiparticles as functions of $\beta=1/T$. We see that the simulation results using $a_n^\dagger$ and $a_n$ in position space agree with the analytical calculations [eq.~\eqref{eq:fp_int}] using the quasiparticle partition functions. Compared to the large mass limit, we also see a relatively large modification on the thermal distribution for the small mass $\bar{m}=0.2$. In fig.~\ref{fig:e_Hint}, we present the energy density as a function of temperature $T$ for the interacting fields in comparison with that for the free fields. Conveniently, we subtracted their respective vacuum energies in the results so that the lowest energy density for both is set to zero. Our simulation agree with the analytical calculations [eq.~\eqref{eq:ene_density_int}] and one can see an increase in energy within the combined system given our choice of parameters. The minor discrepancies are due to insufficiently small trotter steps especially for the large mass limit. Additionally, we also verify the simulation with the exact diagonalization at the smallest trotterization step $\Delta \beta =0.001$ and find them in agreement.

\section{Summary and outlook}\label{sec:summary}

We formulate the quantum field theory for Majorana fermions in 1+1 dimensions on the qubits and study its various thermal properties at finite temperature using quantum simulation algorithm. Specifically, we provide representations of the fermion fields using qubits in both coordinate and momentum space to understand the field theory on a spatial lattice. Using such representations, we carry out quantum computation both analytically and numerically. In our quantum simulations, despite the small number of qubits used in our setup, we showed that the quantum imaginary time evolution (QITE) algorithm~\cite{Motta:2019yya} can be used to study thermal observables such as the energy density and distribution function at finite temperature for the free field theory and the interacting field theory with a homogeneous background. Our numerical results using quantum simulation are compared to analytical calculations and exact diagonalization methods, showing good agreement. This demonstrates the potential of quantum computing for field theory calculations.

Our work is an important first step to understand thermal fixed points using quantum simulation, which can be extended to scalar field theories and gauge theories as subsequent works. The quantum simulation approach is general, naturally avoiding the usual complications associated with non-zero chemical potentials~\cite{Czajka:2021yll} and topological terms~\cite{Honda:2021aum}. Our qubit representation and quantum simulation essentially provide thermal states towards which a quantum fermion system evolves over time during the thermalization process. This prepares for a broader study of quantum field theory in real-time dynamics.

\acknowledgments
We are grateful to Nestor Armesto, Jo$\tilde{a}$o Barata, Meijian Li,  Javier Mas, Alfonso V. Ramallo, Carlos A. Salgado, and Fanyi Zhao for their helpful and valuable discussions. We acknowledge the use of IBM Quantum services for this work. This work is supported by the European Research Council under project ERC-2018-ADG-835105 YoctoLHC; by the Spanish Research State Agency under project PID2020-119632GB-I00; by Xunta de Galicia (Centro singular de investigacion de Galicia accreditation 2019-2022), and by European Union ERDF. W.Q. is also supported by the Marie Sklodowska-Curie Actions Postdoctoral Fellowships under Grant No. 101109293. B.W. acknowledges the support of the Ram\'{o}n y Cajal program with the Grant No. RYC2021-032271-I and the support of Xunta de Galicia under the ED431F 2023/10 project. 

\bibliographystyle{JHEP}
\bibliography{main.bib}
\end{document}